\def\unit#1{\mathord{\thinspace\rm #1}}
\def\dint{\mathop{\displaystyle \int}}
\def\doint{\mathop{\displaystyle \oint}}

\documentclass[12pt,a4paper]{article}
\usepackage{amsmath}
\usepackage{graphicx}
\usepackage{mathpazo}
\usepackage{hyperref}
\usepackage{multimedia}
\usepackage[small,nohug,heads=vee]{diagrams}
\diagramstyle[labelstyle=\scriptstyle]

\usepackage{fancyhdr}
\renewcommand{\maketitle}{
    \begin{center}
      \Large
        {\bf Electrostatics in Stueckelberg-Horwitz-Piron Electrodynamics}
        \vskip .3 true cm
      \small
        Martin Land \\
        \vskip .3 true cm
        Department of Computer Science \\
        Hadassah College \\
        37 HaNevi'im Street, Jerusalem \\
email: martin@hadassah.ac.il
      \end{center}
      \vskip .5 true cm
}
\input{tcilatex}
\begin{document}

\title{}
\author{}
\maketitle



%


%
\begin{abstract}
In this paper, we study fundamental aspects of electrostatics as a special case in 
Stueckelberg-Horwitz electromagnetic theory. 
In this theory, spacetime events $x^\mu(\tau)$ evolve in an unconstrained 8-dimensional phase space,
interacting through five $\tau$-dependent gauge fields induced 
by the current densities associated with their evolutions.
The chronological time $\tau$ was introduced 
as an independent evolution parameter in order to free the laboratory clock $x^0$ to 
evolve alternately 'forward' and 'backward' in time according to the sign of the energy,
thus providing a classical implementation of the Feynman-Stueckelberg interpretation
of pair creation/annihilation.
The resulting theory differs in its underlying mechanics from conventional electromagnetism, but 
coincides with Maxwell theory in an equilibrium limit.

After a brief review of Stueckelberg-Horwitz electrodynamics,
we obtain the field produced by an event in uniform motion and verify that it satisfies
the field equations.
We study this field in the rest frame of the event, where it depends 
explicitly on coordinate time $x^0$ and
the parameter $\tau$, as well as spatial distance $R$.  
Calculating with this generalized Coulomb field, we demonstrate how Gauss's 
theorem and Stoke's theorem apply in 4D spacetime, 
and obtain the fields associated with a charged line and a charged sheet.
Finally, we use the field of the charged sheet to study a static event 
in the vicinity of a potential barrier.  In
all of these cases, we observe a small transfer of mass from the field to the
particle.  It is seen that for an event in the field of an oppositely charged
sheet of sufficient density, the event can reverse time direction, providing a specific model for
pair phenomena.

\end{abstract}
%


\baselineskip7mm \parindent=0cm \parskip=10pt

\section{Introduction}

As described by Born and Wolf, Maxwell's equations were not immediately
accepted as a general theory of classical electrodynamics \cite{B-W_p.xxv-xxvi}.
Maxwell's 1864 formulation summarized all prior research in electricity and
magnetism, including Cavendish's 1771 - 1773 experiments in electrostatics,
Coulomb's characterization of electric and magnetic forces in the 1780s and Faraday's
investigations of time-varying fields in the 1830s.
Nevertheless, this mathematically concise description encountered resistance until its prediction
of electromagnetic waves traveling at the speed of light was verified by Hertz in 1888.
Ultimately, the successful incorporation of optics into electrodynamics through Maxwell's
equations inspired Einstein's 1906 study of the spacetime symmetries underlying the theory
and Fock's 1929 association of potential theory with gauge symmetry in
the quantum mechanics of charged particles \cite{fock_gauge}.
Since that time, 
these symmetry considerations have taken the lead role in extending classical and quantum field theory.  
Thus, the Standard Model of strong and electroweak interactions, which was historically pieced together
from often {\em ad hoc} models of particle behavior, is often presented in contemporary pedagogical 
treatments as a general implementation of {\em a priori} principles of relativistic and gauge invariance.  

An important stage in the development of the Standard Model was the observation by Stueckelberg
\cite{Stueckelberg} that the classical Maxwell theory did not implement all Lorentz symmetries
apparent in relativistic quantum mechanics, and in particular could not provide a classical
account of pair creation/annihilation processes.  
Stueckelberg proposed a reformulation of the classical Lorentz force in which a particle worldline
is traced out dynamically by an evolving spacetime event, and pair processes may be described by a
single event evolving forward and backward though time.
Since the coordinate time is not single-valued in this framework, Stueckelberg introduced 
a monotonically increasing Poincar\'{e} invariant evolution parameter $\tau$, which plays a role
similar to Newtonian time in nonrelativistic mechanics.  
In subsequent work on QED, Stueckelberg's approach was adopted by Feynman \cite{Feynman}
and Schwinger \cite{Schwinger}, who employed an invariant time but de-emphasized its status. 
The parameterized canonical formalism was extended by Horwitz and Piron \cite{H-P}  
to the relativistic classical and quantum mechanics of many particles with interactions.  
Consistency of this approach requires that the gauge symmetry include the parameter
\cite{saad, beyond}, leading to a $\tau$-dependent electrodynamics derived from five gauge fields.  
Although the resulting theory coincides with Maxwell theory in its equilibrium limit, 
it differs in its underlying mechanics from conventional electromagnetism, 
in particular that the total mass of particles and fields is
conserved, but not the masses of individual interacting particles \cite{emlf}.  
By overcoming the
mass-shell constraint, timelike events may evolve continuously through the spacelike region on their
way to reversed-time timelike evolution, and the theory thus provides a solution to a fundamental difficulty
facing Stueckelberg's classical description of pair processes. Because the formalism contains a
4-vector potential and a fifth scalar potential, it 
provides a framework for relativistic generalizations of the classical central force problems
\cite{I, II, scattering, zeeman, stark}.  
Possible experimental signatures have been found in classical and quantum scattering
\cite{larry, letter}. Nevertheless, relatively little
work has been done in the areas of classical electrostatics and
electrodynamics that first led to the Maxwell theory.  

In this paper, we study aspects of laboratory electrostatic phenomenology as a limiting case of 
Stueckelberg-Horwitz electrodynamics. This study differs from Maxwell electrostatics in two essential ways:
first, reversing the direction of  18$^\text{th}$ and 19$^\text{th}$ century research, we
present a theory constructed as the implementation of symmetry principles and seek 
consequences that may lead to experimental observation. 
Second, although we borrow the notion of electrostatics as the study of events
in their rest frames, only spatial coordinates remain fixed.  The advance (or retreat) of time is
an explicit form of motion for the events described by this theory, and 
these events may undergo dynamical forces in this dimension
that affect their measurable characteristics but not their position.
We begin by reviewing the essential features of Stueckelberg-Horwitz electrodynamics in section 2.
In section 3 we obtain the field produced by an event in uniform motion and verify that it satisfies
the field equations. 
In section 4 we specify the field of a
uniformly moving event to the observations in the event's rest frame, which expresses the
generalization of the Coulomb field.  This field is explicitly dependent on coordinate time $x^0$ and
the parameter $\tau$, as well as spatial distance $R$.  
Applying this Coulomb field, we show how Gauss's theorem and Stoke's theorem apply in 4D spacetime, 
and also obtain the fields associated with a charged line and a charged sheet.
In section 5 we treat the field of the charged sheet as an external force acting on an event in
order to formulate the Lorentz force equations for an event in the vicinity of a potential barrier.  In
particular, we study the behavior of a `static' event (held fixed in space) in this field and
demonstrate that its time acceleration produces a small transfer of mass from the field to the
particle.  It is seen that for an event in the field of an oppositely charged
sheet of sufficient density, the event can reverse time direction, providing a specific model for
pair phenomena.

\section{Stueckelberg off-shell electrodynamics}

In seeking a classical description of pair creation/annihilation as a single
worldline generated dynamically by the evolution of an event $x^{\mu }\left(
\tau \right) $, Stueckelberg proposed \cite{Stueckelberg} a generalized
Lorentz force of the form
\begin{equation}
M\left(\ddot{x}^{\mu }+\Gamma _{\nu \rho }^{\mu }\dot{x}^{\nu }
\dot{x}^{\rho}\right) = eF^{\mu \nu }g_{\nu \rho }\dot{x}^{\rho } + K^{\mu }
\mbox{\qquad} \ddot{x}^{\mu }= \dfrac{d^{2}x^{\mu }}{d\tau ^{2}} \mbox{\qquad} \dot{x}^{\mu }=\dfrac{dx^{\mu }}{d\tau }
\label{st-1}
\end{equation}
with metric $g_{\mu \nu }(x)$ and connection $\Gamma _{\nu \rho }^{\mu }$.
The electromagnetic field includes the familiar tensor part $F^{\mu \nu }\left( x\right)$,
as well as a vector part $ K^{\mu }\left( x\right) $ 
whose role is to overcome the mass-shell constraint
\vspace{8pt}
\begin{equation*}
\mbox{\qquad} K^{\mu }=0 \; \Rightarrow \; \left\{ 
\begin{array}{l}
g_{\nu \rho }\dot{x}^{\nu }\dot{x}^{\rho }=\text{constant} \vspace{8pt}\\ 
\text{Timelike event remains timelike} 
%
\end{array}
\right.
\vspace{-04pt}
\end{equation*}

that prevents the event from smoothly entering the spacelike region and thus precludes pair processes.
But finding no physical justification for introducing the vector field $K^{\mu } \ne 0$,
Stueckelberg turned to $\tau $-parameterized
covariant canonical quantum mechanics in flat space with $g_{\mu \nu } \rightarrow \eta_{\mu \nu } = \text{diag}(-1,1,1,1)$
\begin{equation}
i\partial _{\tau }\psi \left( x,\tau \right) =\frac{1}{2m}\left[ p^{\mu
}-eA^{\mu }\left( x\right) \right] \left[ p_{\mu }-eA_{\mu }\left( x\right) 
\right] \psi \left( x,\tau \right)  
\label{K-Stk}
\end{equation}
which permits an event to tunnel from the timelike to spacelike region.
This quantum theory 
enjoys the standard U(1) gauge invariance under local transformations of the type 
\begin{eqnarray}
\psi (x,\tau ) &\longrightarrow &\exp \left[ ie\Lambda (x)\right] \ \psi
(x,\tau )  
\label{100} \\
A_{\mu } &\longrightarrow &A_{\mu }+\partial _{\mu }\Lambda (x).  
\label{110}
\end{eqnarray}
The global gauge invariance associated with this gauge symmetry
is the 
conserved current 
\begin{equation}
\partial _{\mu }j^{\mu }+\partial _{\tau }\rho =0  
\label{140}
\end{equation}
where 
\begin{equation}
\rho =\Bigl|\psi (x,\tau )\Bigr|^{2}\qquad j^{\mu }=-\frac{i}{2M}\Bigl\{\psi
^{\ast }(\partial ^{\mu }-ieA^{\mu })\psi -\psi (\partial ^{\mu }+ieA^{\mu
})\psi ^{\ast }\Bigr\}\ .  
\label{150}
\end{equation}

Stueckelberg \cite{Stueckelberg} regarded (\ref{150}) as a true current,
\smallskip leading to the interpretation of $\Bigl|\psi (x,\tau )\Bigr|^{2}$
as the probability density at $\tau $ of finding the event at the spacetime
point $x$. However, under this interpretation, the non-zero divergence of
the four-vector current $j^{\mu }(x,\tau )$ prevents its identification as
the source of the field $A^{\mu }(x)$. As a remedy, Stueckelberg observed that
assuming $\rho \rightarrow 0$ pointwise as $\tau \rightarrow \pm \infty $,
integration of (\ref{150}) over $\tau $ leads to 
\begin{equation}
\partial _{\mu }J^{\mu }=0\qquad \text{where}\qquad J^{\mu
}(x)=\int_{-\infty }^{\infty }d\tau \;j^{\mu }(x,\tau )\;.  
\label{160}
\end{equation}
However, in the resulting dynamical picture, the fields $A^{\mu }(x)$ that
mediate particle interaction instantaneously at $\tau $ are induced by
currents $J^{\mu }(x)$ whose support covers the particle worldlines, past
and future. There is no {\em a priori} assurance that the particles moving in
these Maxwell fields will trace out precisely the worldlines that induce
the fields responsible for their motion.

In order to obtain a well-posed theory, Sa'ad, Horwitz, and Arshansky \cite{saad}
generalized (\ref{K-Stk}) by introducing a $\tau $-dependent gauge field and
a fifth gauge compensation field. Writing $x^{5}=\tau $ and adopting
the index convention 
\begin{equation}
\alpha ,\beta ,\gamma =0,1,2,3,5  \qquad \qquad
\lambda ,\mu ,\nu =0,1,2,3\qquad \qquad 
i,j,k = 1,2,3
\label{220}
\end{equation}
the Stueckelberg-Schrodinger equation
\begin{equation}
\left[ i\partial _{\tau }+e_{0}a_{5}\left( x,\tau \right) \right] \psi
\left( x,\tau \right) =\frac{1}{2M}\left[ p^{\mu }-e_{0}a^{\mu }\left(
x,\tau \right) \right] \left[ p_{\mu }-e_{0}a_{\mu }\left( x,\tau \right) 
\right] \psi \left( x,\tau \right)  
\label{St-Sh}
\end{equation}
becomes locally gauge invariant under $\tau $-dependent
gauge transformations of the type 
\begin{equation}
\psi \rightarrow e^{ie_{0}\Lambda \left( x,\tau \right) }\psi \mbox{\qquad}
a_{\mu }\rightarrow a_{\mu }+\partial _{\mu }\Lambda \left( x,\tau \right) 
\mbox{\qquad}a_{5}\rightarrow a_{5}+\partial _{\tau }\Lambda \left( x,\tau
\right) .
\end{equation}
Writing the classical Lagrangian \cite{emlf} as
\begin{equation}
L=\dot{x}^{\mu }p_{\mu }-K=\tfrac{1}{2}M\dot{x}^{\mu }\dot{x}_{\mu }+e_{0} \dot{x}^{\alpha }a_{\alpha }
\label{lag}
\end{equation}
the Lorentz force
\begin{equation}
\dfrac{d}{d\tau }\dfrac{\partial L}{\partial \dot{x}_{\mu }}-\dfrac{\partial L}{\partial x_{\mu }}=0 \quad \longrightarrow \quad
\dfrac{d}{d\tau } \Big[ M\dot{x}^{\mu } + e_{0} a^{\mu } (x,\tau )\Big] = e_{0} \dot{x}^{\alpha }\partial^\mu a_{\alpha }(x,\tau )
\end{equation}
takes the form
\begin{equation}
M\ddot{x}^{\mu } = e_{0} f^\mu_{\; \;  \alpha} (x,\tau )\dot{x}^\alpha
= e_{0} \big[ f^\mu_{\; \;  \nu} (x,\tau )\dot{x}^\nu + f^\mu_{\; \;  5} (x,\tau ) \big] 
\label{lf}
\end{equation}
where
\begin{equation}
f^\mu_{\; \;  \alpha} = \partial^\mu a_\alpha - \partial_\alpha a^\mu \mbox{\qquad\qquad} \dot{x}^5
= \dot{\tau} = 1
\label{f-def}
\end{equation}
and Stueckelberg's vector field --- now $\tau$-dependent --- may be identified as the term
\begin{equation}
K^\mu = f^\mu_{\; \;  5} (x,\tau ) = 
\partial^\mu a_5  (x,\tau )- \partial_\tau a^\mu (x,\tau ).
\label{lf5}
\end{equation}
In this formalism, the mass shell constraint is overcome by the exchange of mass between 
particles and fields seen in 
\begin{equation}
\frac{d}{d\tau }(-\frac{1}{2}M\dot{x}^{2})=-M\dot{x}^{\mu }\ddot{x}_{\mu
}=-e_{0}\ \dot{x}^{\mu }(f_{\mu 5}+f_{\mu \nu }\dot{x}^{\nu })=-e_{0}\ \dot{x
}^{\mu }f_{\mu 5} 
\label{mass}
\end{equation}
but total mass-energy-momentum of particle and fields is conserved \cite{emlf, ald}.
This relaxation of the mass-shell constraint breaks general
reparameterization invariance, but under the boundary conditions 
\begin{equation}
\tau \rightarrow \pm \infty \qquad \Rightarrow
\qquad a^5\left( x,\tau \right) \rightarrow 0 \qquad \qquad j^{5}\left( x,\tau \right)\rightarrow 0
\label{bc}
\end{equation}
the remaining $\tau $-translation symmetry is associated, via
Noether's theorem, with dynamic conservation of mass. 
Since the gauge fields possess a non-zero
mass spectrum, this theory has been called off-shell electrodynamics.

The electromagnetic field $f_{\alpha \beta }\left( x,\tau \right) $ is made
a dynamical quantity by adding a kinetic term to the action \cite{saad}. In analogy to
standard Maxwell theory, one may adopt the formal designation $f^{\mu
5}=\eta ^{55}f_{~~5}^{\mu }=-f_{~~5}^{\mu }$ and choose the form
\begin{equation}
-\frac{\lambda }{4}f^{\alpha \beta }\left( x,\tau \right) f_{\alpha \beta
}\left( x,\tau \right)
\label{field-kin}
\end{equation}
leading to the classical action
\begin{equation}
S = \int d\tau \; \dfrac{1}{2}M\dot{x}^{\mu }\dot{x}_{\mu }
+ \int d^4z d\tau \left\{ e_0 a_\alpha (z,\tau) j^\alpha (z,\tau)
-\frac{\lambda }{4}f_{\alpha \beta } (z,\tau)f^{\alpha \beta }(z,\tau)
\right\}
\end{equation}
with locally conserved five-component event current
\begin{equation}
j^\alpha (z,\tau) = \dot{x}^\alpha(\tau) \delta^4\left(z-x(\tau)\right) .
\label{20}
\end{equation}
The homogeneous field equation
\begin{equation}
\epsilon ^{\alpha \beta
\gamma \delta \epsilon }\partial _{\alpha }f_{\beta \gamma }=0
\label{homo}
\end{equation}
follows automatically from the definition (\ref{f-def}) of the fields $f^{\alpha \beta }$. 
The inhomogeneous field equations are obtained as
\begin{equation}
\partial _{\beta }f^{\alpha \beta }\left( x,\tau \right) =\frac{e_{0}}{
\lambda }j^{\alpha }\left( x,\tau \right) =ej^{\alpha }\left( x,\tau \right)
=e\dot{z}^{\alpha }\left( \tau \right) \delta ^{4}\left[ x-z\left( \tau
\right) \right]  
\label{pM} 
\end{equation}
by variation of the action.
Under the boundary conditions (\ref{bc}) the 
standard Maxwell theory is extracted as the equilibrium limit of (\ref{homo}) and
(\ref{pM}) by integration over the worldline
\begin{equation}
\left. 
\begin{array}{c}
\partial _{\beta }f^{\alpha \beta }\left( x,\tau \right) =ej^{\alpha }\left(
x,\tau \right) \\ 
\\ 
\partial _{\lbrack \alpha }f_{\beta \gamma ]}=0
\end{array}
\right\} \underset{\int d\tau }{\mbox{\quad}\xrightarrow{\hspace*{1cm}}
\mbox{\quad}}\left\{ 
\begin{array}{c}
\partial _{\nu }F^{\mu \nu }\left( x\right) =eJ^{\mu }\left( x\right) \\ 
\\ 
\partial _{\lbrack \mu }F_{\nu \rho ]}=0
\end{array}
\right.
\end{equation}
where 
\begin{equation}
F^{\mu \nu }(x)=\int_{-\infty }^{\infty }d\tau \; f^{\mu \nu }\left( x,\tau
\right) \mbox{\qquad}A^{\mu }(x)=\int_{-\infty }^{\infty } d\tau \; a^{\mu
}\left( x,\tau \right)  . 
\label{c-cat-2}
\end{equation}
This integration has been called concatenation \cite{concat} and as in (\ref{160}) links the
event current $j^{\mu }\left( x,\tau \right) $ with the particle current
$ J^{\mu }(x)$ defined on the entire particle worldline, which by (\ref{20}) recovers the standard
covariant expression. 
It is seen from (\ref{St-Sh}) and (\ref{c-cat-2}) that $e_{0}$ and 
$\lambda $ must have dimensions of time, so that the dimensionless ratio
$ e=e_{0}/\lambda $ can be identified as the Maxwell charge.
The microscopic $\tau$-dependent fields have been called pre-Maxwell fields.

The wave equation derived from (\ref{pM}) is 
\begin{equation}
\partial _{\alpha }\partial ^{\alpha }a^{\beta }\left( x,\tau \right)
=\left( \partial _{\mu }\partial ^{\mu }-\;\partial _{\tau }^{2}\right)
a^{\beta }\left( x,\tau \right) =-ej^{\beta }\left( x,\tau \right)
\label{wave}
\end{equation}
and the principal part Green's function \cite{green} found from 
\begin{equation}
\left( \partial _{\mu }\partial ^{\mu }-\;\partial _{\tau }^{2}\right)
G\left( x,\tau \right) =-\delta ^{5}\left( x,\tau \right)
\end{equation}
is
\begin{equation}
G\left( x,\tau \right) =-{\frac{1}{{2\pi }}}\delta (x^{2})\delta \left( \tau
\right) -{\frac{1}{{2\pi ^{2}}}\frac{\partial }{{\partial {x^{2}}}}}\ {\frac{
{\theta (x^{2}-\tau ^{2})}}{\sqrt{{x^{2}-\tau ^{2}}}}}=D\left( x\right)
\delta (\tau )-G_{correlation}\left( x,\tau \right) .  
\label{grn}
\end{equation}
The first term has support on the lightcone at instantaneous $\tau $, and
recovers the standard Maxwell Green's function under concatenation. The
second term has spacelike support (${x^{2}>\tau ^{2}\geq 0}$) and vanishes
under concatenation, so it may contribute to correlations but not to Maxwell
potentials.  Terms of this type have been studied in \cite{jigal}. 
The conserved current (\ref{140}), electromagnetic action (\ref{field-kin}) and wave equation
(\ref{wave}) suggest an underlying five-dimensional symmetry --- O(3,2) or O(4,1) depending on the
sign of $\eta^{55}$ --- however the equal-$\tau$ part of (\ref{grn}) breaks this
symmetry to a vector-plus-scalar
representation of O(3,1).  Stueckelberg-Horwitz electrodynamics may be called a 5D theory of five
gauge fields, as a shorthand for the 4+1 implementation of O(3,1) Lorentz symmetry.
The equal-$\tau$ part is seen to satisfy
\begin{equation}
\left( \partial _{\mu }\partial ^{\mu }-\;\partial _{\tau }^{2}\right)
D\left( x\right) \delta (\tau ) = \delta^4\left( x\right) \delta (\tau ) 
- D\left( x\right) \delta^{\prime\prime} (\tau )
\end{equation}
and may be treated as providing 
solutions the wave equation by neglecting
terms associated with $\delta ^{\prime \prime}(\tau )$.  An integrated study of the relation
between the two pieces of the Green's function is
forthcoming.

Because a static event --- one whose position is space remains unchanged --- evolves at a
timelike velocity along the $t$-axis in its rest frame, pre-Maxwell theory does not contain any precise
equivalent of the motionless charge that produces the Coulomb force in Maxwell electrostatics.
Low energy Coulomb scattering is studied by solving the Lorentz force (\ref{lf}) in the field of
a spatially static event  
\begin{equation}
x\left(\tau\right) =\left(t, {\mathbf x} \right) = \left( \tau ,{\mathbf 0}\right) \quad \longrightarrow\quad  
\left\{ 
\begin{array}{l}
j^0\left( x,\tau \right) = j^5\left( x,\tau \right) = \delta(t-\tau) \, \delta^3(x) \\ 
\\ 
{\bf j} \left( x,\tau \right)= 0
\end{array}
\right.
\end{equation}
where boldface signifies 3D spatial quantities.
Applying (\ref{grn}) to this current leads to
\begin{equation}
a^{0}\left( x,\tau \right) = a^{5}\left( x,\tau \right)
={\frac{e}{{4\pi }}}\frac{\delta \left( x^{0}-\vert {\mathbf x} \vert -\tau \right) }{\vert {\mathbf x} \vert }
\end{equation}
which recovers the correct Coulomb potential under concatenation 
\begin{equation}
A^{0}(x)=\int d\tau \, a^{0}\left( x,\tau \right) = {\frac{e}{{4\pi R}}}
\end{equation}
but incorrectly describes the microscopic dynamics
\begin{equation}
M\mathbf{\ddot{x}}=-e_{0}\nabla \left[ a^{0}\left( x,\tau \right)
+  a^{5}\left( x,\tau \right) \right] =
-\frac{e_{0}}{4\pi} \nabla \, \frac{\delta ( x^{0}-\vert {\mathbf x} \vert -\tau ) }{\vert {\mathbf
x} \vert } .
\end{equation}
Since the $\tau$-translation invariance of the theory leaves
the Coulomb interaction invariant under
shift of the worldline origin
$x\left( \tau \right) =\left( \tau ,{\mathbf 0}\right) \rightarrow
\left( \tau +\delta\tau ,{\mathbf 0}\right)$,
it was suggested in \cite{larry} to relax the synchronization of
interacting events by taking the induced current to be
\begin{equation}
j^{\alpha }\left( x,\tau \right) \longrightarrow j_{\varphi }^{\alpha
}\left( x,\tau \right) =\int ds\ \varphi \left( \tau
-s\right) j^{\alpha }\left( x,s\right) \mbox{\qquad}\varphi (\tau )=\frac{1}{
2\lambda }e^{-|\tau |/\lambda }
\label{fi-cur}
\end{equation}
leaving the concatenated current unchanged
\begin{equation}
J^{\mu }\left( x\right) =\int d\tau \ j_{\varphi }^{\mu
}\left( x,\tau \right) =\int ds~d\tau \ \varphi \left(
\tau -s\right) j^{\mu }\left( x,s\right) =\int ds\ j^{\mu }\left( x,s\right) . 
\end{equation}
This modification leads to a Yukawa potential with reasonable low energy limit 
\begin{equation}
M\mathbf{\ddot{x}}=-e_{0}\nabla \left[ a^{0}\left( x,\tau \right)
+a^{5}\left( x,\tau \right) \right] 
=-2e_{0}\nabla a^{0}\left( x,\tau \right) =e^{2}\nabla \left[ 
\frac{e^{-\left\vert \mathbf{x}\right\vert /\lambda }}{{4\pi }\left\vert 
\mathbf{x}\right\vert }\right] .
\end{equation}
The source $j_{\varphi }^{\alpha }\left( x,\tau \right) $
for the pre-Maxwell field is interpreted as a smoothed current density induced
by an ensemble of events $x^{\alpha }\left( \tau +\delta \tau \right) $ along
a particle worldline, where $\delta \tau $ is given by the normalized
distribution $\varphi \left( \tau \right) $. 
The distribution $\varphi \left( \tau \right) $ provides a cutoff for the
photon mass spectrum, which we take to be the conventional experimental
uncertainty in photon mass ($\Delta m_{\gamma }\simeq 10^{-17}\unit{eV}$ 
\cite{pdg}), leading to a value of about 400 seconds for $\lambda $. The
limit $\lambda \rightarrow 0$ restores $\varphi \left( \tau \right)
\rightarrow \delta \left( \tau \right) $ and the limit $\lambda \rightarrow
\infty $ restores standard Maxwell theory. Since the form of $\varphi \left(
\tau \right) $ given in (\ref{fi-cur}) represents the distribution of
interarrival times of events in a Poisson-distributed stochastic process,
this choice suggests an information-theoretic interpretation for the
underlying the relationship between the current density and the ensemble of
events from which it is induced.

The smoothed current can be introduced through the action \cite{high-order},
by adding a higher $\tau $-derivative term to the electromagnetic part. The
substitution 
\begin{equation}
S_{em}\rightarrow \int d^{4}xd\tau \left[ e_{0}j^{\alpha }a_{\alpha }-\frac{
\lambda }{4}f^{\alpha \beta }\left( x,\tau \right) f_{\alpha \beta }\left(
x,\tau \right) -\frac{\lambda ^{3}}{4}\left[ \partial _{\tau }f^{\alpha
\beta }\left( x,\tau \right) \right] \left[ \partial _{\tau }f_{\alpha \beta
}\left( x,\tau \right) \right] \right]
\label{smoothed}
\end{equation}
preserves Lorentz and gauge invariance, and leaves the action first order in
spacetime derivatives. Defining a field interaction kernel 
\begin{equation}
\Phi \left( \tau \right) =\delta \left( \tau \right) -\lambda ^{2}\delta
^{\prime \prime }\left( \tau \right) =\frac{1}{2\pi }\int d\kappa \,\left[
1+\left( \lambda \kappa \right) ^{2}\right] \,e^{-i\kappa \tau }
\end{equation}
which is seen from
\begin{equation}
\int_{-\infty }^{\infty }ds~\Phi (\tau -s)\varphi (s)=\delta (\tau
)\rightarrow \varphi \left( \tau \right) =\int \frac{d\kappa }{2\pi }\frac{
e^{-i\kappa \tau }}{1+\left( \lambda \kappa \right) ^{2}}=\frac{1}{2\lambda }
e^{-|\tau |/\lambda }  
\label{inverse}
\end{equation}
to be the inverse function to $\varphi (\tau )$, the action becomes
\begin{equation}
S_{em}=\int d^{4}xd\tau ~e_{0}j^{\alpha }a_{\alpha }-\frac{\lambda }{4}\int
d^{4}x\,d\tau \,ds\ f^{\alpha \beta }\left( x,\tau \right) \Phi (\tau
-s)f_{\alpha \beta }\left( x,s\right) .  
\label{Action-2}
\end{equation}
The Euler-Lagrange equations 
\begin{equation}
\partial _{\beta }f_{\Phi }^{\alpha \beta }(x,\tau )=\partial _{\beta }\int
ds\,\Phi (\tau -s)f^{\alpha \beta }(x,s)=ej^{\alpha }\left( x,\tau \right)
\label{EuM}
\end{equation}
can be inverted to recover
\begin{equation}
\partial _{\beta }f^{\alpha \beta }\left( x,\tau \right) =ej_{\varphi
}^{\alpha }\left( x,\tau \right) =e\int ds~\varphi \left( \tau -s\right)
j^{\alpha }\left( x,s\right) 
\label{pMm}
\end{equation}
using (\ref{inverse}). The action (\ref{Action-2}), in which the statistical synchronization
performed by $\Phi (\tau -s)$ is made explicit, has the advantage of
permitting the usual study of symmetries and being amenable to second
quantization, where the factor $\left[ 1+\left( \lambda \kappa \right) ^{2}
\right] ^{-1}$ provides a natural mass cutoff for the off-shell photon that
renders off-shell quantum field theory super-renormalizable at two-loop
order \cite{high-order}.

\section{Induced fields}

\subsection{Li\'{e}nard-Wiechert potential}

The potential and field strength induced by the motion of an arbitrary event were obtained in
\cite{ald} in studying the self-interaction problem in the context of Stueckelberg-Horwitz
electrodynamics.
Writing the smoothed current for an arbitrary event $r^{\alpha }\left( \tau \right) $ as
\begin{equation}
j_{\varphi }^{\alpha }\left( x,\tau \right) = c \int ds~\varphi \left( \tau-s\right) 
\dot{r}^{\alpha }\left( s\right) \delta ^{4}\left[ x-r\left(
s\right) \right]
\label{s-curr}
\end{equation}
the Li\'{e}nard-Wiechert potential found from the Maxwell part of the Green's function (\ref{grn}) is 
\begin{eqnarray}
a^{\alpha }\left( x,\tau \right) &&\mbox{\hspace{-20pt}}=\dfrac{e}{2\pi c}\int
d^{4}x^{\prime }d\tau ^{\prime }
\delta \left( \left( x-x^{\prime } \right) ^{2}\right)\theta^{ret}
\delta \left(
\tau -\tau ^{\prime }\right) j_{\varphi }^{\alpha }\left( x^{\prime },\tau^{\prime } \right) \notag \\
&&\mbox{\hspace{-20pt}}=\frac{e}{2\pi }\int ds~\varphi \left( \tau
-s\right) \dot{r}^{\alpha }\left( s\right) \delta \left( \left( x-r\left(s\right) \right) ^{2}\right) \theta ^{ret}
\label{LWf}
\end{eqnarray}
where $\theta^{ret}$ imposes retarded $x^0$ causality and we write the speed of light $c$
explicitly. Using the identity
\begin{equation}
\dint d\tau f\left( \tau \right) \delta \left[
g\left( \tau \right) \right] =\dfrac{f\left( \tau_R\right) }{\left\vert g^{\prime
}\left( \tau_R\right) \right\vert }\mbox{\quad},\mbox{\quad}\tau_R=g^{-1}\left( 0\right)
\label{delta-id}
\end{equation}
we obtain
\begin{equation}
a^{\beta }\left( x,\tau \right) =\frac{e}{4\pi }\varphi \left( \tau -\tau_R\right) \frac{
\dot{r}^{\beta }\left( \tau_R\right) }{\left( x^{\mu }-r^{\mu }\left( \tau_R\right)
\right) \dot{r}_{\mu }\left( \tau_R\right)}
\label{LWu}
\end{equation}
where the retarded time $\tau_R$ satisfies $[x-r(\tau_R)]^2 =0$ and
$ \theta ^{ret}=\theta \left( x^{0}-r^{0}\left( \tau_R\right) \right) =1$.

It is convenient to express the field quantities as elements of a Clifford algebra
with basis vectors
\begin{equation}
e_{\alpha }\cdot e_{\beta }= \eta_{\alpha \beta } = \text{diag} \left( -1,1,1,1,-1 \right)
\qquad\qquad
e_{\alpha } \wedge e_{\beta } = e_{\alpha } \otimes e_{\beta } - e_{\beta } \otimes e_{\alpha }
\end{equation}
and Clifford product
\begin{equation}
e_{\alpha } e_{\beta }=e_{\alpha }\cdot e_{\beta } +e_{\alpha }\wedge e_{\beta } 
\end{equation}
where we refer to the index convention (\ref{220}).  Separating spacetime and scalar quantities as
\begin{eqnarray}
r(\tau) = r^\mu (\tau) e_\mu 
\qquad\qquad
r^5 = c \tau \\
d = \partial_\mu e^\mu 
\qquad\qquad
\partial_5 = \dfrac{1}{c} \partial_\tau 
\end{eqnarray}
the field strength tensors 
\begin{equation}
f = \dfrac{1}{2} \; f^{\mu\nu} \; e_\mu \wedge e_\nu \qquad\qquad
f^{5} = f^{5\mu} \; e_5 \wedge e_\mu = e_{5}\wedge \epsilon  .
\end{equation}
are expressed as Clifford bivectors and (\ref{f-def}) takes the form
\begin{equation}
f = d \wedge a \qquad \qquad
\epsilon  = - \partial_5 a - d a^5 .
\end{equation}
The covariant equivalent of a spatially static charge is a uniformly evolving event
\begin{equation}
\mbox{\qquad}r\left(\tau\right)= u \tau = \left(u^{0}\tau,\mathbf{u}\tau\right) 
\end{equation}
with constant timelike velocity $\dot{r} = u$, which in its rest frame 
simply advances along the time axis as $t=(u^0 / c) \tau$.
The induced potential is found from (\ref{LWu}) to be
\begin{equation}
a(x,\tau )={\frac{e}{{4\pi }}} \frac{u \varphi (\tau - \tau_R)}{\left\vert u\cdot z\right\vert }
\mbox{\qquad\quad} a^5(x,\tau )={\frac{e}{{4\pi }}} \frac{c \varphi (\tau - \tau_R)}{\left\vert u\cdot z\right\vert }
\label{pots}
\end{equation}
along the line of observation
\begin{equation}
z=x-r(\tau_R) = x - u\tau_R\mbox{\qquad} \mbox{\qquad} z^2=0.
\end{equation}
Writing the timelike velocity $u$ in terms of the unit vector
$\hat{u}$ 
\begin{equation}
u^{2} <0\mbox{\qquad\qquad}u=\left\vert u\right\vert \hat{u}\mbox{\qquad\qquad}\hat{u}
^{2}=-1\mbox{\qquad\qquad}u^{2}=-\left\vert u\right\vert ^{2}
\end{equation}
the observation line $z$ can be separated into components
\begin{equation}
z_{\parallel }=-\hat{u}\left( \hat{u}\cdot
z\right) \mbox{\qquad\qquad}z_{\perp }=z+\hat{u}\left( \hat{u}\cdot z\right)
\end{equation}
which satisfy 
\begin{eqnarray}
& z_{\parallel }^{2} =\hat{u}^{2}\left( \hat{u}\cdot z\right) ^{2}=-\left( 
\hat{u}\cdot z\right) ^{2} & 
\label{gm1}
\\
& z_{\perp }^{2} =z^{2}+2\left( 
\hat{u}\cdot z\right) ^{2}-\left( \hat{u}\cdot z\right) ^{2}
=\left( \hat{u}\cdot z\right) ^{2}=-z_{\parallel }^{2} & 
\label{gm2}
\\
& \left( u\cdot z\right) ^{2} =\left\vert u\right\vert ^{2}\left( \hat{u}\cdot z\right)
^{2}=-\left\vert u\right\vert ^{2}z_{\parallel }^{2} . &
\label{gm3}
\end{eqnarray}%
The condition of retarded causality 
\begin{equation}
z^{2} =\tau_R^{2}u^{2}-2\tau_Ru\cdot x+x^{2} = 0
\end{equation}
relates the field to the location of the event along the backward lightcone of the observation
point.  This implicit choice of $\tau_R$ and its gradient
\begin{equation}
0 =  d (z^2) = 2 \left(\tau_R d\tau_R u^{2} - \tau_R u -d\tau_R u\cdot x +x \right)
= 2 \left[- \left( u\cdot z\right) d \tau_R+z\right]
\label{kin1}
\end{equation}
define the essential kinematics of the system as embodied in the following expressions:
\begin{equation}
d\tau_R = \dfrac{z}{u\cdot z} \qquad
\left( u \cdot d \right) \tau_R =\frac{u \cdot z}{u\cdot z}=1 \mbox{\qquad} 
\left( z \cdot d \right) \tau_R =\frac{z^2}{u\cdot z}=0
\label{kin2}
\end{equation}
\begin{equation}
d \left( u\cdot z\right) = d \left( u\cdot
x-u^{2}\tau_R\right)=\frac{\left( u\cdot z\right) u-u^{2}z}{u\cdot z}
=\left\vert u\right\vert ^{2}\frac{z_{\perp}}{u\cdot z}
\label{kin3}
\end{equation}
\begin{equation}
d\frac{1}{\left( u\cdot z\right) ^{n}} =\frac{-n\left[ \left( u\cdot
z\right) u-u^{2}z\right] }{\left( u\cdot z\right) ^{n+2}}=
\frac{-n\left\vert u^{2}\right\vert z_{\perp }}{\left( u\cdot z\right) ^{n+2}
}
\label{kin4}
\end{equation}
\begin{equation}
d\cdot z = d\cdot \left( x-u\tau_R\right) =d\cdot x-u\cdot d\tau_R=3
\label{kin5}
\end{equation}
\begin{equation}
d\wedge z = d\wedge  \left( x-u\tau_R\right) =  d\tau_R \wedge u = \frac{u\wedge z}{u\cdot z}
\label{kin6}
\end{equation}
%
\begin{equation}
d\wedge \hat{z}=d\wedge \frac{z}{\left\vert z\right\vert }=-\frac{1}{
\left\vert z\right\vert }\frac{z\wedge u}{u\cdot z}-\hat{z}\wedge \frac{z}{
\left\vert z\right\vert ^{2}}=\frac{u \wedge \hat{z} }{u\cdot z}
\label{kin9}
\end{equation}
We will refer to equations (\ref{gm1}) -- (\ref{gm3}) and (\ref{kin2}) -- (\ref{kin9}) collectively as the
kinematic equations.

\subsection{Field strengths}

To find the field strengths we must calculate field derivatives of the type
$\partial ^{\alpha }a^{\beta }(x,\tau )$. 
As in the Maxwell case, the spacetime derivative is most conveniently 
found by returning to (\ref{LWf}) with $\dot{r} = u$ to write
\begin{equation}
\partial ^{\mu }a^\beta(x,\tau )
={\frac{e}{{2\pi }}}\int d\tau_R\
\varphi (\tau -\tau_R) \ u^\beta \ \partial ^{\mu }\delta 
\left[ \left( x-r \left( \tau_R\right) \right) ^{2}\right] 
\end{equation}
and combining the expressions
\begin{equation}
\partial ^{\mu }\delta \left[ \left( x-r \left( \tau_R\right) \right) ^{2}
\right] 
=2\delta ^{\prime }\left[ \left( x-r \left( \tau_R\right) \right)
^{2}\right] \left( x^{\mu }-r ^{\mu }\left( \tau_R\right) \right)
\end{equation}
%
%
%
%
%
%
\begin{equation}
\frac{d}{d\tau_R}\delta \left[ \left( x-r \left( \tau_R\right) \right) ^{2}\right]
=-2\delta ^{\prime }\left[
\left( x-r \left( \tau_R\right) \right) ^{2}\right] \ u \ \cdot \left( x-r \left( \tau_R\right) \right) 
\end{equation}
to obtain
\begin{equation}
\partial ^{\mu }a^\beta(x,\tau )=-{\frac{e}{{2\pi }}}\int d\tau_R\ \varphi
(\tau -\tau_R) u^\beta \frac{x^{\mu }-r ^{\mu }\left(
\tau_R\right) }{u \cdot \left( x-r \left( \tau_R\right)
\right) }\frac{d}{d\tau_R}\delta \left[ \left( x-r \left( \tau_R\right) \right) ^{2}
\right] .
\end{equation}
Integrating by parts
%
%
%
%
%
\begin{equation}
\partial ^{\mu }a^\beta(x,\tau )={\frac{e}{{2\pi }}}\int d\tau_R\ \left[ \frac{
d}{d\tau_R}\varphi (\tau -\tau_R)  u^\beta  \frac{x^{\mu }-r
^{\mu }\left( \tau_R\right) }{  u  \cdot \left( x-r \left(
\tau_R\right) \right) }\right] \delta \left[ \left( x-r \left( \tau_R\right) \right)
^{2}\right]
\end{equation}
and again using identity (\ref{delta-id})
%
%
%
%
we find
%
%
%
%
%
\begin{equation}
%
f = d \wedge a(x,\tau ) = {\frac{e}{{4\pi }}}\frac{1}{\left\vert u\cdot z\right\vert }\frac{d}{d\tau_R}
 \left[ \varphi (\tau -\tau_R)\frac{z \wedge u}{u\cdot z}\right] 
\label{f-der}
\end{equation}
for the spacetime components and
\begin{equation}
%
d a^5(x,\tau ) = 
{\frac{e}{{4\pi }}}\frac{1}{\left\vert u\cdot z\right\vert }\frac{d}{d\tau_R}
 \left[ \varphi (\tau -\tau_R)\frac{zu^5}{u\cdot z}\right]
\label{a5-der}
\end{equation}
for the fifth gauge field.
Since (\ref{LWu}) shows that $a(x,\tau )$ depends on $\tau$ only through $\varphi \left( \tau
-\tau_R\right)$, 
we may write
\begin{equation}
\partial^5 a = -\dfrac{1}{c}\partial_\tau a(x,\tau )= -{\frac{e
}{{4\pi }}}\varphi ^{\prime }(\tau -\tau_R)\frac{u}{\left\vert u\cdot
z\right\vert } \qquad \qquad \varphi ^{\prime }(\tau ) = \dfrac{d}{d\tau} \varphi(\tau)
\label{p5}
\end{equation}
for the $\tau$-derivative term required in $f^5$.
%
%
%
%
Using (\ref{p5}) and applying
\begin{equation}
\dfrac{d}{d\tau_R} u \wedge z = u \wedge \dfrac{d}{d\tau_R} (x-u\tau_R) = - u \wedge u = 0
\end{equation}
\begin{equation}
\frac{d}{d\tau_R} u \cdot z = - u \cdot (x-u\tau_R) = -u^2
\end{equation}
\begin{equation}
\frac{d}{d\tau_R} \dfrac{z}{u \cdot z} = \dfrac{-u(u \cdot z) + zu^2}{(u \cdot z)^2}
= -u^2 \dfrac{z_\perp}{(u \cdot z)^2}
\end{equation}
to (\ref{f-der}) and (\ref{a5-der})
we are finally led to
\begin{equation}
f = -{\frac{e}{{4\pi }}}\left[ \varphi \left( \tau -\tau_R\right) \frac{\left(
z\wedge u\right) u^{2}}{\left( u\cdot z\right) ^{3}}-\varphi ^{\prime
}\left( \tau -\tau_R\right) \frac{z\wedge u }{\left( u\cdot
z\right) ^{2}}\right]
\label{78}
\end{equation}
and
\begin{equation}
f^{5} = e_{5}\wedge {\frac{ec}{{4\pi }}}\left[ \varphi \left( \tau
-\tau_R\right) \frac{u^{2}z_{\perp } }{\left( u\cdot z\right) ^{3}}
-\varphi ^{\prime }\left( \tau -\tau_R\right) \frac{z-u\left( u\cdot
z\right) /c^{2}}{\left( u\cdot z\right) ^{2}}\right] 
=  e_{5}\wedge \epsilon \; .
\label{79}
\end{equation}
Notice that integration of (\ref{78}) over $\tau$ involves
\begin{equation}
\int d\tau \; \varphi ( \tau ) = 1 \qquad \qquad
\int d\tau \; \varphi ^{\prime }( \tau ) = \varphi ( \infty ) - \varphi ( -\infty ) =
0
\label{int_phi}
\end{equation}
so that taking $u^2 = -c^2$ under concatenation recovers 
\begin{equation}
F(x)  = {\frac{e}{{4\pi }}}  \frac{\left(
z\wedge u\right) c^{2}}{\left( u\cdot z\right) ^{3}} 
%
%
\end{equation}
which is the standard Maxwell field for a uniformly moving particle.

\subsection{Pre-Maxwell equations}

Separating the four-vector and fifth scalar components, the pre-Maxwell equations 
(\ref{homo}) and (\ref{pM}) can be written in 4D component form as
\begin{equation} 
\begin{array}{lcl}
\partial _{\nu }\;f^{\mu \nu }-\partial _{\tau }\;f^{5\mu }=ej^{\mu } _{\varphi} 
& \mbox{\qquad} &\partial _{\mu }\;f^{5\mu }=e\rho _{\varphi} \vspace{8pt}\\ 
\partial _{\mu }f_{\nu \rho }+\partial _{\nu }f_{\rho \mu }+\partial _{\rho }f_{\mu \nu }=0 
& &\partial _{\nu }f_{5\mu }-\partial _{\mu }f_{5\nu }+\partial _{\tau }f_{\mu \nu }=0 \\
\end{array}
\end{equation}
which exposes the analogy with 3-vector Maxwell equations
\begin{equation} 
\begin{array}{lcl}
\nabla \times \mathbf{B}-\partial _{0}\mathbf{E}=e\mathbf{J} 
& \mbox{\qquad} \mbox{\qquad} & \nabla \cdot \mathbf{E}=eJ^{0} \vspace{8pt}\\ 
\nabla \cdot \mathbf{B}=0
& &\nabla \times \mathbf{E}+\partial _{0}\mathbf{B}=0  \\
\end{array}
\end{equation}
by observing that the field $f^{5\mu} = \epsilon^\mu$ plays a role analogous to the Maxwell electric
field $F^{0i} = E^i$ 
while $f^{\mu\nu}$ plays the role of the magnetic field $F^{ij} = \varepsilon^{ijk}B_k$. 
In the index-free notation, with $c$ written explicitly, the pre-Maxwell equations take the form
\begin{equation} 
\begin{array}{lcl}
\partial _{\mu }\;f^{5\mu }=e\rho &\mbox{\quad} \longrightarrow \mbox{\quad}& 
d \cdot \epsilon = \dfrac{e}{c} j^5_{\varphi } = e\rho_{\varphi } \rule[-024pt]{0cm}{024pt}\\ 
\partial _{\nu }\;f^{\mu \nu }-\partial _{\tau }\;f^{5\mu }=ej^{\mu } & \mbox{\quad}
\longrightarrow \mbox{\quad}&
-d\cdot f-\dfrac{1}{c}\partial _{\tau }\epsilon =\dfrac{e}{c}j_{\varphi }\rule[-024pt]{0cm}{024pt}\\ 
\partial _{\mu }f_{\nu \rho }+\partial _{\nu }f_{\rho \mu }+\partial _{\rho }f_{\mu \nu }=0 
& \longrightarrow & d\wedge f = 0 \rule[-024pt]{0cm}{024pt}\\
\partial _{\nu }f_{5\mu }-\partial _{\mu }f_{5\nu }+\partial _{\tau }f_{\mu \nu }=0 
&\mbox{\quad} \longrightarrow \mbox{\quad}& d\wedge \epsilon +\dfrac{1}{c}\partial _{\tau }f=0\\
\end{array}
\end{equation}
which must apply to the solution for the uniformly moving event.


To verify the Gauss law we calculate the 4-divergence
\begin{equation}
d \cdot \epsilon = d \cdot {\frac{ec}{{4\pi }}}\left[ \varphi \left( \tau
-\tau_R\right) \frac{u^{2}z_{\perp } }{\left( u\cdot z\right) ^{3}}
-\varphi ^{\prime }\left( \tau -\tau_R\right) \frac{z-u\left( u\cdot
z\right) /c^{2}}{\left( u\cdot z\right) ^{2}}\right]
\end{equation}
using
\begin{equation}
d \varphi \left( \tau - \tau_R\right) =  - \varphi^{\prime } \left( \tau - \tau_R\right)
d\tau_R \; .
\end{equation}
From the kinematic equations we find
\begin{eqnarray}
%
d\cdot \left( \varphi  \frac{u^2 z_{\perp }}{\left( u\cdot
z\right) ^{3}}\right) &=&- \varphi ^{\prime }\frac{u^2z_{\perp }^{2}}{\left(
u\cdot z\right) ^{4}}+\varphi \frac{u^2d\cdot z_{\perp }}{\left( u\cdot
z\right) ^{3}}+\varphi \frac{-3u^2\left\vert
u^{2}\right\vert z_{\perp }^{2}}{\left( u\cdot z\right) ^{5}} \notag \\
&=&-\varphi ^{\prime }\frac{u^2 z_{\perp }^{2}}{\left(
u\cdot z\right) ^{4}}+3\varphi \left( \frac{u^2 \left(
u\cdot z\right) ^{2}}{\left( u\cdot z\right) ^{5}}-\frac{u^2 \left\vert
u^{2}\right\vert z_{\perp }^{2}}{\left( u\cdot z\right) ^{5}}\right)\notag \\
&=&-\varphi ^{\prime } \frac{u^2 z_{\perp }^{2}}{\left(
u\cdot z\right) ^{4}}
\end{eqnarray}%
%
%
%
%
%
for the first term, and  
\begin{equation}
d\cdot \left( \varphi ^{\prime } \ \frac{z-u\left( u\cdot
z\right) /c^{2}}{\left( u\cdot z\right) ^{2}}\right) =-\varphi ^{\prime }\frac{%
\left\vert u\right\vert ^{2}z_{\parallel }^{2}}{\left( u\cdot z\right) ^{4}}
-\frac{1}{c^{2}}%
\varphi ^{\prime \prime } \frac{1}{ u\cdot z }
\end{equation}%
for the second term.
Since $u^2 z_{\perp }^{2}= \vert u^2 \vert z_{\parallel }^{2}$ we are led to
\begin{equation}
d\cdot \epsilon = -\varphi ^{\prime \prime } \frac{1}{c^{2}\left( u\cdot z\right) } \sim 0 
\label{gauss}
\end{equation}
where the second derivative $\varphi ^{\prime \prime }$ is associated with the term $\delta^{\prime
\prime }(\tau)$ in (\ref{grn}), which we take to be equivalent to zero.
Thus, Gauss's law is verified in the source-free region.  We return to the integral form of Gauss's law in
section 4 to establish equality including the enclosed source.


The time derivative of $\epsilon$ is easily found to be
\begin{equation}
\dfrac{1}{c} \dfrac{\partial}{\partial \tau} \epsilon =  {\frac{e}{{4\pi }}}\left[ \varphi ^{\prime }\left( \tau
-\tau_R\right) \frac{u^{2}z_{\perp } }{\left( u\cdot z\right) ^{3}}
-\varphi ^{\prime \prime }\left( \tau -\tau_R\right) \frac{z-u\left( u\cdot
z\right) /c^{2}}{\left( u\cdot z\right) ^{2}}\right] .
\label{df5}
\end{equation}
The divergence of the field $f$ requires
\begin{equation}
d\cdot f=-{\frac{e}{{4\pi }}}d\cdot \left[ \varphi  
\frac{\left( z\wedge u\right) u^{2}}{\left( u\cdot z\right) ^{3}}-\varphi
^{\prime } \frac{\left( z\wedge u\right) \left( u\cdot
z\right) }{\left( u\cdot z\right) ^{3}}\right]
\end{equation}
from which the first term leads to 
\begin{eqnarray}
d\cdot \left[ \varphi  \frac{\left( z\wedge u\right)
u^{2}}{\left( u\cdot z\right) ^{3}}\right] &=&d\varphi 
\cdot \frac{\left( z\wedge u\right) u^{2}}{\left( u\cdot z\right) ^{3}}%
+\varphi  \frac{d\cdot \left( z\wedge u\right) u^{2}}{%
\left( u\cdot z\right) ^{3}}+\varphi  u^{2}\left[ d%
\frac{1}{\left( u\cdot z\right) ^{3}}\right] \cdot \left( z\wedge u\right)
\notag
\\
&=&\varphi ^{\prime } \frac{u^{2}z}{\left( u\cdot
z\right) ^{3}}
\label{df1}
\end{eqnarray}
and the second term is
\begin{eqnarray}
d\cdot \left[ \varphi ^{\prime } \frac{\left( z\wedge
u\right) \left( u\cdot z\right) }{\left( u\cdot z\right) ^{3}}\right]
&=&d\varphi ^{\prime } \cdot \frac{\left( z\wedge
u\right) \left( u\cdot z\right) }{\left( u\cdot z\right) ^{3}}+\varphi
^{\prime } d\cdot \frac{\left( u\cdot z\right) \left(
z\wedge u\right) }{\left( u\cdot z\right) ^{3}} \notag \\
&=&-{\frac{e}{{4\pi }}}\left[ \varphi ^{\prime }\left( \tau
-s\right) \frac{u^{2}z_{\perp } }{\left( u\cdot z\right) ^{3}}%
-\varphi ^{\prime \prime }\left( \tau -s\right) \frac{\left( u\cdot z\right)
z}{\left( u\cdot z\right) ^{3}}\right] .
\label{df2}
\end{eqnarray}
Combining (\ref{df1}) and (\ref{df2}) we find
\begin{equation}
d \cdot f = - {\frac{e}{{4\pi }}}\left[ \varphi ^{\prime }\left( \tau
-\tau_r\right) \frac{u^{2}z_{\perp } }{\left( u\cdot z\right) ^{3}}
-\varphi ^{\prime \prime }\left( \tau -\tau_r\right) \frac{z-u\left( u\cdot
z\right) /c^{2}}{\left( u\cdot z\right) ^{2}}\right]
\label{df4}
\end{equation}
which when compared with (\ref{df5}) verifies Amp\`ere's law in the source-free region.

The exterior derivative of $f$
\begin{equation}
d\wedge f=-{\frac{e}{{4\pi }}}d\wedge \left[ \varphi \left( \tau -s\right) 
\frac{\left( z\wedge u\right) u^{2}}{\left( u\cdot z\right) ^{3}}-\varphi
^{\prime }\left( \tau -s\right) \frac{ z\wedge u  }{\left( u\cdot z\right) ^{2}}\right]
\end{equation}
produces three types of term:
\begin{equation}
d \varphi \wedge \left( z\wedge u\right) =  - \varphi^{\prime } 
\dfrac{z}{u \cdot z}\wedge \left( z\wedge u\right) = 0
\end{equation}
%
%
\begin{equation}
d\wedge \left( z\wedge u\right) = \left( d\wedge z\right)\wedge u =0
\end{equation}
%
%
\begin{equation}
\left[ d\frac{1}{\left( u\cdot z\right) ^{n}}\right] \wedge \left( z\wedge
u\right) =\frac{-n\left\vert u^{2}\right\vert z_{\perp }}{\left( u\cdot
z\right) ^{n+2}}\wedge \left( z_{\perp }\wedge u\right) =0
\end{equation}
and thus we recover the absence of electromagnetic monopoles in the form
\begin{equation}
d\wedge f = 0 .
\end{equation}
In the pre-Maxwell theory, the 4-divergence of the field $\varepsilon^\mu (x,\tau)$ locates the 
event density $\rho(x,\tau)$ as its source. The Maxwell field $F^{ij}$ is induced by motion
$J^i(x)$ of the charge density $J^0(x)$ and has no monopole source ---
in the pre-Maxwell theory the field $f^{\mu\nu}$ is induced by the motion $j^\mu$ of the 
event density $\rho(x,\tau)$ and has no monopole source.


The time derivative of the $f$ is easily found to be
\begin{equation}
\dfrac{1}{c} \dfrac{\partial}{\partial \tau} f = 
-{\frac{e}{{4\pi }}}\left[ \varphi ^{\prime } \frac{%
\left( z\wedge u\right) u^{2}}{\left( u\cdot z\right) ^{3}}-
\varphi ^{\prime \prime } \frac{ z\wedge u  }{\left( u\cdot z\right) ^{2}}%
\right] .
\label{dtauf5}
\end{equation}
Writing the exterior derivative of $\epsilon$ in the form
\begin{equation}
d\wedge \epsilon ={\frac{ec}{{4\pi }}}d\wedge \left[ \varphi  \frac{%
\left\vert u\right\vert ^{2}z_{\perp }}{\left( u\cdot z\right) ^{3}}-\varphi
^{\prime } \frac{z}{\left( u\cdot z\right) ^{2}}
+\varphi
^{\prime } \frac{u }{c^{2}\left( u\cdot z\right)}
\right]
\label{97}
\end{equation}
three terms contribute
\begin{equation}
d\wedge \varphi  \frac{\left\vert u\right\vert ^{2}z_{\perp }}{\left( u\cdot z\right) ^{3}}
=d\varphi  \wedge \frac{\left\vert u\right\vert ^{2}z_{\perp }}{\left( u\cdot z\right) ^{3}}
+\varphi  \ d  \wedge \frac{\left\vert u\right\vert ^{2}z_{\perp }}{\left( u\cdot z\right) ^{3}}
=-\varphi ^{\prime } \frac{z\wedge u}{\left( u\cdot
z\right) ^{3}}
\end{equation}
\begin{equation}
d\wedge \varphi ^{\prime } \frac{z}{\left( u\cdot z\right) ^{2}}
=d\varphi ^{\prime } \wedge \frac{z}{ \left( u\cdot z\right) ^{2}}+
\varphi ^{\prime } d\wedge \frac{z}{\left( u\cdot z\right) ^{2}}
= \varphi ^{\prime } \frac{z\wedge u}{\left( u\cdot z\right) ^{3}} 
\end{equation}
\begin{equation}
d\wedge \varphi ^{\prime } \frac{u}{
c^{2}\left( u\cdot z\right) } = d \varphi ^{\prime }
\wedge \frac{u}{c^{2}\left( u\cdot z\right) }+\varphi
^{\prime } d\frac{1}{\left( u\cdot z\right) }\wedge 
\frac{u}{c^{2}}
=\varphi ^{\prime } \frac{u^{2}z\wedge u}{
c^{2}\left( u\cdot z\right) ^{3}}-\varphi ^{\prime \prime }
\frac{z\wedge u}{c^{2}\left( u\cdot z\right) ^{2}}
\end{equation}
leading to
\begin{equation}
d\wedge \epsilon = {\frac{e}{{4\pi }c}}\left[ \varphi ^{\prime }
\frac{\left(z\wedge u\right)u^{2}}{\left( u\cdot z\right) ^{3}}-\varphi ^{\prime \prime }
\frac{z\wedge u}{\left( u\cdot z\right) ^{2}}\right] . 
\label{102}
\end{equation}
Comparing (\ref{97}) and (\ref{102}) we recover 
\begin{equation}
d\wedge \epsilon +\frac{1}{c}\dfrac{\partial}{\partial \tau} \; f=0 
\end{equation}
which we recognize as Faraday's Law.
This confirms that the fields (\ref{78}) and (\ref{79}) are an essentially kinematic
solution of the pre-Maxwell equations.

\section{Electrostatics}

In order to compare the phenomenology of 3D electrostatics with the standard Maxwell theory we
describe the source event in its rest frame, so that the event 
evolves uniformly along its time axis.  We take the event velocity to be 
\begin{equation}
u = (c,0) = ce_{0}\mbox{\qquad}u^{2}=-c^{2}
\end{equation}
and write the point of observation as
\begin{equation}
x= (ct,\mathbf{x}) = (ct,R\mathbf{\hat{x}}) 
\end{equation}
so that the line of observation satisfies
\begin{equation}
z^2 = (x-ce_{0}\tau_R)^2 = 0 \quad \rightarrow \quad \tau_R = t-\frac{R}{c}
\quad \rightarrow \quad z\left( \tau_R\right) =R\left( e_{0}+\mathbf{\hat{x}} \right)
\end{equation}
\begin{equation}
z_\perp = R\mathbf{\hat{x}} \mbox{\qquad\quad} u \cdot z = -cR
\mbox{\qquad\quad} u \wedge z = cR \left( e_{0} \wedge \mathbf{\hat{x}} \right).
\end{equation}
In this frame (\ref{pots}) becomes
\begin{equation}
a^0(x,\tau )={\frac{e}{{4\pi }}} \frac{\varphi (\tau - t+R/c)}{R}
\mbox{\qquad\quad}
{\mathbf a}(x,\tau )=0
\mbox{\qquad\quad}
a^5(x,\tau )=a^0(x,\tau )
\label{potss}
\end{equation}
so that
\begin{equation}
f = d \wedge a
= e_0 \wedge \left(- \nabla a^0 \right)
\qquad \qquad
\epsilon  = \partial^5 a - d a^5
= -\left( e_0 \partial_5 a^0+ d a^5\right) 
\label{f109}
\end{equation}
and we see that in addition to the 4-gradient of the scalar potential $a^5$, 
the $x^0$ evolution contributes motion terms $a^0$ to both the magnetic-type $f$ field strength
and the electric-type field $\epsilon$. 
This situation differs structurally from Maxwell electrostatics
\vspace{12pt}
\begin{equation}
\left. 
\begin{array}{c}
A^0(x )={\dfrac{e}{{4\pi R}}} \\ 
\\ 
{\mathbf A}(x)=0
\end{array}
\right\} {\mbox{\quad}\xrightarrow{\hspace*{1cm}}
\mbox{\quad}}\left\{ 
\begin{array}{c}
{\mathbf E}(x)= -\nabla A^0 \\ 
\\ 
{\mathbf B}(x)=0
\end{array}
\right.
\label{A-field}
\end{equation}

in which the field is pure electric and derives entirely from the
gradient of the ``scalar'' potential $A^0$.  Separating the 
gradient into space and time components $d = \left( \partial_0 , \nabla \right)$
we find from (\ref{f109}) 
\begin{equation}
\boldsymbol{\epsilon }=f^{5i}\boldsymbol{e}_i = -\nabla a^5
\mbox{\qquad\quad} \epsilon^0=f^{50}
= \partial^5 a^0 -\partial^0 a^5  = -\dfrac{2}{c} \; \partial_\tau a^5 
\label{curlless}
\end{equation}
so that the space part of $\epsilon$ has zero curl.

\subsection{Coulomb law}

From the potentials (\ref{potss}) or directly from (\ref{78}) and (\ref{79}) we are led to 
\begin{equation}
\epsilon ^{0} =-{\frac{e}{{2\pi}}}\dfrac{\varphi ^{\prime }\left( \tau -t+R/c\right)}{cR}
\qquad
\boldsymbol{\epsilon } = {\frac{e}{{4\pi }}}\left[ \frac{\varphi \left( \tau
-t+R/c\right) }{R^{2}}-\frac{\varphi ^{\prime }\left( \tau -t+R/c\right) }{cR
}\right] \mathbf{\hat{x}}
\label{107}
\end{equation}
and
\begin{equation}
f =   {\frac{e}{{4\pi }}}\left[ \frac{\varphi \left( \tau
-t+R/c\right) }{R^{2}}-\frac{\varphi ^{\prime }\left( \tau -t+R/c\right) }{cR
}\right] e_{0} \wedge\mathbf{\hat{x}}
\label{108}
\end{equation}
so that in terms of the spatial components $e_{i}=f^{0i}$ and $b_{i}=\epsilon _{ijk}f^{jk}$ we find
$\mathbf{e} = \boldsymbol{\epsilon }$ and $\mathbf{b} = 0$.  Using (\ref{int_phi}) we see that the
concatentated electric field is
\begin{equation}
\mathbf{E} =   \frac{e }{4\pi R^{2}} \hat{x}
\end{equation}
recovering the standard Coulomb law.  From (\ref{lf}) we may write the Lorentz force 
experienced by a test event at rest, that is $(\dot{x}^0, \mathbf{\dot{x}}) = (c,{\mathbf 0}) $, as
\begin{eqnarray}
M~\ddot{t}&=&\frac{e_0}{c^{2}} \; \mathbf{e}\cdot \mathbf{\dot{x}}+\frac{e_0}{c } \; \epsilon ^{0} 
= \varepsilon \left( \tau -t+R/c\right)\frac{e^2}{4 \pi \lambda R c }
e^{-\left\vert \tau -t+R/c \right\vert /\lambda }
\\ 
M~\mathbf{\ddot{x}}&=&e_0 \; \mathbf{e} \; \dot{t}+\frac{e_0}{c}
\; \mathbf{\dot{x}} \; {\mathbf \times} \; \mathbf{b}+e_0 \; \boldsymbol{\epsilon }
= \frac{e^2}{4 \pi R^2 }
e^{-\left\vert \tau -t+R/c \right\vert /\lambda }\mathbf{\hat{x}}
\end{eqnarray}
where we used
\begin{equation}
\varphi ^{\prime }\left( \tau \right) =\frac{1}{2\lambda }\frac{d}{d\tau }%
e^{-\left\vert \tau \right\vert /\lambda }=-\varepsilon \left( \tau \right) 
\frac{1}{2\lambda ^{2}}e^{-\left\vert \tau \right\vert /\lambda } \qquad \qquad 
\varepsilon \left( \tau \right) = \text{signum} (\tau) .
%
\end{equation}%
Placing the test charge at $t = \tau+R/c + \alpha$ with $\vert \alpha \vert \ll \lambda$,
that is, along the forward
lightcone of the event producing the field, the force becomes
\begin{equation}
M~\ddot{t} = \varepsilon \left( \alpha\right)\frac{e^2}{4 \pi \lambda R c }
\qquad \qquad 
M~\mathbf{\ddot{x}}= \frac{e^2}{4 \pi R^2 }\mathbf{\hat{x}}
%
\end{equation}
and we see that the temporal acceleration for this configuration depends on the sign of $\alpha$.  In
the absence of some mechanism to fix $\dot x^0$, a test event slightly behind the 
lightcone with $\alpha < 0$ will be further
decelerated on the time axis until the temporal separation weakens the force.  Similarly, a test event with
$\alpha > 0$ will be further accelerated on the time axis.  For a test event at rest in space, 
these possible changes in $\dot x^0$ --- the energy of the test event --- imply a change of mass,
which will be further studied in section 5.

Applying the $x^0$-derivative and 3-gradient to (\ref{107}) 
we find
\begin{equation}
\frac{1}{c}\dfrac{\partial}{\partial t} \epsilon ^{0} ={\frac{
e}{{2\pi c}^{2}R}}\varphi ^{\prime \prime
}\left( \tau -t+R/c\right)
\end{equation}
\begin{equation}
\nabla \cdot \boldsymbol{\epsilon }={\frac{e}{{4\pi }}}\left[ 4\pi \varphi
\left( \tau -t+R/c\right) \delta ^{3}\left( \mathbf{x}\right) -\frac{1}{
c^{2}R}\cdot \varphi ^{\prime \prime }\left( \tau -t+R/c\right) \right]
\end{equation}
which combine to confirm Gauss's law in differential form 
\begin{equation}
\frac{1}{c}\dfrac{\partial}{\partial t} \epsilon ^{0} + \nabla \cdot \boldsymbol{\epsilon }=
e\varphi \left( \tau -t+R/c\right) \delta ^{3}\left( \mathbf{x}\right) 
+ {\frac{e}{{4\pi c}^{2}R}}\varphi ^{\prime \prime
}\left( \tau -t+R/c\right)
= e\rho_{\varphi }(x)
\label{gauss-3d}
\end{equation}
where as in (\ref{gauss}) we take $\varphi ^{\prime \prime} \sim 0$.  To obtain the $\delta^3(x)$
term, we used
\begin{equation}
\nabla \cdot \dfrac{\mathbf{\hat{x}} \;}{R^2} = 4\pi \delta^3(\mathbf{x}) 
\label{grad-1overR}
\end{equation}
which is proven by applying Gauss's theorem to the 3D volume integral of the right hand side. 
In 4D the integral form of Gauss's law (\ref{gauss-3d}) can be found by performing spacetime integral
\begin{equation}
\int \partial _{\mu }\epsilon ^{\mu }~d^{4}x = e \int \rho_{\varphi }(x) \; d^{4}x =e
\label{gauss1}
\end{equation}
over the volume formed 
as the product of the time axis and a 3-sphere in space. To demonstrate the method, we consider
the standard Coulomb field for a point charge and directly calculate
\begin{equation}
\int \nabla \cdot \mathbf{E}~d^{3}x=\int \nabla \cdot \dfrac{\mathbf{\hat{x}}}{4\pi R^2}~d^{3}x
=e\int \delta ^{3}\left( \mathbf{x}\right) \mathbf{~}d^{3}x=e
\end{equation}
to obtain the total charge.  
Instead of the usual application of Gauss's theorem in spherical coordinates, we
express the field in cylindrical coordinates
\begin{equation}
\mathbf{E}(\rho,\phi,z)=\frac{e\left( \rho \mathbf{\hat{n}}+z\mathbf{\hat{z}}\right) }{
4\pi \left( \rho ^{2}+z^{2}\right) ^{3/2}}
\mbox{\qquad}\mbox{\qquad}  \mathbf{\hat{n}}
=\left( \cos \phi ,\sin \phi \right) \mbox{\qquad} \rho = \sqrt{x^2+y^2}
\end{equation}
and enclose the charge in a long thin cylinder.  Neglecting the flux through the
vanishingly small ends of the cylinder, we take the surface element to be
\begin{equation}
d\mathbf{S}=\mathbf{\hat{n}}dS=\mathbf{\hat{n}}\rho d\phi dz 
\end{equation}
so that the surface integral is easily evaluated as
\begin{equation}
\int \mathbf{E}\cdot d\mathbf{S} = 
\int_{0}^{2\pi }d\phi \int_{-\infty }^{\infty }dz
\; \rho \; \frac{e\left( \rho \mathbf{\hat{n}}+z\mathbf{
\hat{z}}\right) }{4\pi \left( \rho ^{2}+z^{2}\right) ^{3/2}}\cdot \mathbf{\hat{n}}
=\frac{e\rho ^{2}}{2}\left. \frac{z}{\rho
^{2}\left( \rho ^{2}+z^{2}\right) ^{1/2}}\right\vert _{-\infty }^{\infty }=e.
\end{equation}
For the pre-Maxwell field, (\ref{gauss1}) becomes 
\begin{equation}
e= \int \partial _{\mu }\epsilon ^{\mu }~d^{4}x  = \int \nabla \cdot \boldsymbol{\epsilon}~d^{4}x
\label{gauss2}
\end{equation}
because
$\varphi^{\prime } (\pm \infty) =0$ assures 
\begin{equation}
\int dx^0~\partial_0\epsilon ^{0}=  {\frac{e}{{2\pi }}}\frac{1}{cR}\int
dt~\varphi ^{\prime \prime }\left( \tau -t+R/c\right) =0.
\end{equation}
Again neglecting the flux through the ends and taking the 3D boundary element to be
\begin{equation}
d\mathbf{S}
= \mathbf{\hat{x}} R^{2}~d\Omega ~dt 
\end{equation}
so that using $\varphi (-x) =\varphi (x) $ and $\varphi (\pm \infty) =0$,
we obtain the surface integral
\begin{equation}
\int
\boldsymbol{\epsilon }\cdot d\mathbf{S}
=\int {\frac{e}{{4\pi }}}\left[ \frac{\varphi \left(
\tau -t+R/c\right) }{R^{2}}-\frac{\varphi ^{\prime }\left( \tau
-t+R/c\right) }{cR}\right] \mathbf{\hat{x}}\cdot \mathbf{\hat{x}}
R^{2}~d\Omega ~dt =e
\end{equation}
verifying Gauss's law in integral form for the space-static solution.  In the pre-Maxwell theory,
the Gauss law expresses the equality of the total charged event density contained in 
a region of 4D spacetime with the total flux passing through a 3D boundary surrounding that region.


As in Maxwell theory, the field $\epsilon$ is characterized by its divergence and exterior
derivative.  In 3+1 components, 
%
\begin{equation}
d\wedge \epsilon = -e_{0}\wedge \left( \partial _{0}\boldsymbol{\epsilon }+\nabla \epsilon
^{0}\right) +\nabla \wedge \boldsymbol{\epsilon }
= -e_{0}\wedge \left( \partial _{0}\boldsymbol{\epsilon }+\nabla \epsilon ^{0}\right)
\label{123}
\end{equation}
%
so that from (\ref{107}) we find
\begin{equation}
\partial _{0}\boldsymbol{\epsilon }=-{\frac{e}{{4\pi c}}}\left[ 
\frac{\varphi ^{\prime }\left( \tau -t+R/c\right) }{R^{2}}-\frac{\varphi
^{\prime \prime }\left( \tau -t+R/c\right) }{cR}\right] \mathbf{\hat{x}}
\end{equation}
\begin{equation}
\nabla \epsilon ^{0} ={\frac{e }{{2\pi c}}}\left[ \frac{\varphi
^{\prime }\left( \tau -t+R/c\right) }{R^{2}}-\frac{\varphi ^{\prime \prime
}\left( \tau -t+R/c\right) }{cR}\right] \mathbf{\hat{x}}
\end{equation}
leading to
\begin{equation}
\partial _{0}\boldsymbol{\epsilon }+\nabla \epsilon
^{0}={\frac{e}{{4\pi c}}}\left[ 
\frac{\varphi ^{\prime }\left( \tau -t+R/c\right) }{R^{2}}-\frac{\varphi
^{\prime \prime }\left( \tau -t+R/c\right) }{cR}\right] \mathbf{\hat{x}} .
\label{127}
\end{equation}
Using (\ref{108}) we calculate 
\begin{equation}
\frac{1}{c}\dfrac{\partial}{\partial \tau} \; f = {\frac{e}{{4\pi c}}}\left[ \frac{\varphi
^{\prime }\left( \tau -t+R/c\right) }{R^{2}}-\frac{\varphi
^{\prime \prime }\left( \tau -t+R/c\right) }{cR}\right] e_{0}\wedge \mathbf{\hat{x}} 
\label{128}
\end{equation}
to recover
\begin{equation}
d\wedge \epsilon +\frac{1}{c}\dfrac{\partial}{\partial \tau} \; f=0
\label{faraday}
\end{equation}
as Faraday's law.

In Maxwell electrostatics, Stokes theorem along with (\ref{A-field}) establishes the electric field as
conservative through
\begin{equation}
\doint \mathbf{E} \cdot \mathbf{dl}
= \int \left(\nabla \times \mathbf{E} \right) \cdot \mathbf{dS}
= \int \left(\nabla \times \nabla A^0 \right) \cdot \mathbf{dS} 
= 0 . 
\end{equation} 
This integral is seen to represent work by writing
\begin{equation}
\doint \mathbf{E} \cdot \mathbf{dl}
=\doint \mathbf{E} \cdot \dfrac{\mathbf{dl}}{dt} dt
=\doint \mathbf{E} \cdot \mathbf{v} \; dt
\end{equation} 
and noting that the integrand is the time-derivative of energy in the covariant Lorentz force for
Maxwell's equations.  Similarly writing 
\begin{equation}
\doint \epsilon \cdot dx
=\doint \epsilon \cdot \dot{x} \; d\tau
\end{equation} 
we recognize the integrand as the {\em fifth} Lorentz force equation (\ref{lf5}) so that the
integral represents the total mass change to the event.
Considering the space part of the pre-Maxwell field,
it follows from (\ref{curlless}) that 
\begin{equation}
\doint \boldsymbol{\epsilon} \cdot \mathbf{dl}
= \int \left(\nabla \wedge \boldsymbol{\epsilon} \right) \cdot \mathbf{dS}
= \int \left(\nabla \wedge \nabla a^5 \right) \cdot \mathbf{dS} 
= 0
\end{equation}
for any closed path through space. But for a general closed path in
spacetime, the Faraday law (\ref{faraday}) leads to
\begin{equation}
\doint \epsilon \cdot dl = \int \left( d \wedge \epsilon \right) \cdot dS
= -\frac{1}{c} \dfrac{\partial}{\partial \tau} \int f \cdot dS
\end{equation}
which need not vanish.  For example, representing a surface in spacetime as 
\begin{equation}
dS
= \left( e_{0}\wedge \mathbf{\hat{x}} \right) \ dt \ dR
\end{equation}
and using the Clifford identity
%
\begin{equation}
\left(y \wedge x \right) \cdot \left(a \wedge b \right) = y \cdot \left[x \cdot \left(a \wedge b \right)\right]
= \left(x \cdot a\right) \left(y \cdot b\right) - \left(x \cdot b\right) \left(y \cdot a\right) 
\end{equation}
to find
\begin{equation}
\left( e_{0}\wedge \mathbf{\hat{x}} \right) \cdot \left( e_{0}\wedge \mathbf{\hat{x}} \right) = 1
\end{equation}
we are led from (\ref{108}) to
\begin{equation}
\doint \epsilon \cdot dl 
= -\frac{1}{c} \dfrac{\partial}{\partial \tau} \int {\frac{e}{{4\pi }}}\left[ \frac{\varphi \left( \tau
-t+R/c\right) }{R^{2}}-\frac{\varphi ^{\prime }\left( \tau -t+R/c\right) }{cR
}\right]\ dt \ dR 
\end{equation}
which again using $\varphi^{\prime } (\pm \infty) =0$ 
becomes
\begin{equation}
\doint \epsilon \cdot dl 
 ={\frac{e}{{4\pi c}}}\int 
\frac{\varphi ^{\prime }\left( \tau -t+R/c\right)}{R^2} \ dt \ dR . 
\end{equation}
For a sufficiently long time interval 
\begin{equation}
\int \varphi ^{\prime }\left( \tau -t+R/c\right) \ dt \longrightarrow \varphi\left( - \infty \right) - \varphi\left(  \infty \right) = 0
\end{equation}
and so 
the $\epsilon$ field is seen to be mass-conservative.  But if we consider the short closed, timelike
path illustrated in Figure 1,

$$
\begin{diagram}
 &\mbox{\qquad\quad}\mbox{\qquad\quad}\mbox{\qquad\quad}    &\mbox{\qquad} \left( cT + \frac{R_2-R_1}{u/c},R_2\right)\\
 & \ruTo~{\text{ forward timelike}}& \dTo(2,4)_{\text{ time retreat at } R_2}\\
\left( cT,R_1\right)\mbox{\qquad}  & &\\
&&\\
 & &\mbox{\qquad} \left( -cT+ \frac{R_2-R_1}{u/c},R_2\right)\\
\uTo(2,4)^{\text{time advance at } R_1 }& \ldTo~{\text{ reverse timelike}}& \\
\left( -cT,R_1\right)\mbox{\qquad}  & & \\
\end{diagram}
$$
\begin{center} Figure 1 \end{center}

we find that for $R_2 - R_1 \ll cT$
\begin{eqnarray}
\doint \epsilon \cdot dl &\simeq& {\frac{e}{{4\pi cR_1}}} \left[ \varphi \left( \tau -T+\dfrac{R_1}{c}\right) 
- \varphi \left( \tau +T+\dfrac{R_1}{c}\right) \right] \rule[-12pt]{0cm}{12pt} \notag \\ 
&& \mbox{\qquad} + 
{\frac{e}{{4\pi cR_2}}} \left[\varphi \left( \tau +T+\dfrac{R_2}{c}\right)
- \varphi \left( \tau -T+\dfrac{R_2}{c}\right) \right]
\end{eqnarray}
suggesting that net mass must be invested in moving an event around a closed timelike curve.  
Because this effect depends on the functional form of $\varphi (x)$ and the time parameter
$\lambda$, it would provide an experimental signature for the theory.

\subsection{Line charge}

We now consider a long straight charged line oriented along the $z$-axis,
with charge per unit length $\lambda_e $.  In cylindrical coordinates
\begin{equation}
\mathbf{x}=\left( \boldsymbol{\rho },z\right) \mbox{\qquad}\mbox{\quad} \boldsymbol{\rho }=\left(
x,y\right) =\rho \boldsymbol{\hat{\rho}}\mbox{\qquad}\mbox{\quad} \rho =\sqrt{x^{2}+y^{2}}
\end{equation}
the fields $\epsilon$ and $\mathbf{e}$ are found by replacing $R = \sqrt{\rho^{2}+z^{2}}$ in 
the fields (\ref{107}) and (\ref{108}) and integrating along the line to find
\begin{equation}
\mathbf{e} = \boldsymbol{ \epsilon} ={\frac{\lambda_e}{{4\pi }}}\int dz\left( \frac{\varphi \left( \tau
-t+\frac{\left( \rho ^{2}+z^{2}\right) ^{1/2}}{c}\right) }{\left( \rho
^{2}+z^{2}\right) ^{3/2}}-\frac{\varphi ^{\prime }\left( \tau -t+\frac{
\left( \rho ^{2}+z^{2}\right) ^{1/2}}{c}\right) }{c\left( \rho
^{2}+z^{2}\right) }\right) \left( \rho \boldsymbol{\hat{\rho}},z\right) \\
\label{e-line}
\end{equation}
\begin{equation}
\epsilon^0 = -{\frac{
\lambda_e}{{4\pi }}}\int dz \frac{\varphi ^{\prime
}\left( \tau -t+\frac{\left( \rho ^{2}+z^{2}\right) ^{1/2}}{c}\right) }{
c\left( \rho ^{2}+z^{2}\right) ^{1/2}} .
\end{equation}
We may get a sense of these expressions by taking the sharp distribution $\varphi(x) = \delta (x)$
which permits us to easily carry out the $z$-integration to obtain
\begin{equation}
\mathbf{e} = \boldsymbol{ \epsilon} ={\frac{\lambda_e}{{2\pi }}}\left( \frac{\theta \left( t
-\rho /c-\tau\right)\rho }{c\left[ \left( t-\tau \right) ^{2}-\rho ^{2}/c^{2}\right]
^{3/2}}-\frac{\delta \left( t-\rho /c-\tau \right) }{\sqrt{
\left( t-\tau \right) ^{2}-\rho ^{2}/c^{2}}}\right) \boldsymbol{\hat{\rho} } .
\end{equation}
We observe the retarded causality in the vanishing of the field for $\tau > \tau_R = t -\rho / c$.
Returning to (\ref{e-line}) and using (\ref{int_phi}) to integrate over $\tau$,
the remaining $z$ integration can be readily
performed to obtain the concatenated electric field 
\begin{equation}
{\mathbf E}(x) = \int d\tau \; {\mathbf e}(x,\tau)  ={\frac{\lambda_e}{{4\pi }}} \int dz\frac{1}{\left( \rho ^{2}+z^{2}\right)
^{3/2}}\left( \rho \boldsymbol{\hat{\rho}},z\right)={\frac{\lambda_e}{{2\pi \rho}}}\left(
\boldsymbol{\hat{\rho}},0\right)
\end{equation}
in agreement with the standard expression.

\subsection{Charge Sheet}

We finally consider a charged sheet in the $x-y$ plane with charge per unit area $\sigma $. 
Integrating the
potential in (\ref{potss}) over $x$ and $y$ with $R = \sqrt{x^{2}+y^{2}+z^{2}}$ we find
\begin{equation}
a^0(x,\tau ) = a^5(x,\tau ) =
{\frac{\sigma c}{{4\pi }}}\int dx^{\prime }dy^{\prime } \; 
 \frac{\varphi \left(
\tau -t+\frac{1}{c}\sqrt{\left( x-x^{\prime }\right) ^{2}+\left( y-y^{\prime
}\right) ^{2}+z^{2}}\right)}{c\sqrt{\left( x-x^{\prime }\right)
^{2}+\left( y-y^{\prime }\right) ^{2}+z^{2}}} . 
%
\end{equation}
Changing to radial coordinates $(x,y) \rightarrow (\rho,\theta)$
we obtain
\begin{equation}
a^0(x,\tau ) = a^5(x,\tau ) 
= {\frac{\sigma c}{{4\pi }}}\int d \theta d \rho \; 
\frac{\varphi \left( \tau -t+\frac{1}{c}\sqrt{\rho ^{2}+z^{2}}\right)}{c\sqrt{\rho  ^{2}+z^{2}}}
\end{equation}
which by change of variable $\zeta = \frac{1}{c} \sqrt{\rho ^{2}+z^{2}}$ becomes
\begin{equation}
a^0(x,\tau ) = a^5(x,\tau ) 
={\frac{\sigma c}{{2}}}\int_{\left\vert
z\right\vert /c}^{\infty }\ \varphi \left( \tau -t+\zeta \right) d\zeta .
\end{equation}
We calculate the fields from (\ref{curlless}) and using
$\varphi \left( \tau -t+ \infty \right) = 0$ to find
\begin{equation}
\epsilon ^{0} =-\sigma 
\int_{\left\vert z\right\vert /c}^{\infty }\ \partial _{\zeta }\varphi \left(
\tau -t+\zeta \right) d\zeta
=\sigma \varphi \left( \tau -t+\frac{\left\vert z\right\vert }{c}\right)
\label{ep0-sheet}
\end{equation}
and
\begin{equation}
\boldsymbol{\epsilon }= -{\frac{\sigma c}{{2}}}\nabla
\int_{\left\vert z\right\vert /c}^{\infty }\ \varphi \left( \tau -t+\zeta
\right) d\zeta  ={\frac{\sigma }{{2}}}\varphi \left( \tau -t+\frac{\left\vert
z\right\vert }{c}\right) \nabla \left\vert z\right\vert
\end{equation}
so that
\begin{equation}
\boldsymbol{\epsilon }(x,\tau) = {\mathbf e}(x,\tau) = {\frac{\sigma }{{2}}}\varepsilon(z)\varphi \left( \tau -t+\frac{\left\vert
z\right\vert }{c}\right) \; \mathbf {\hat{z}}  
\label{e-sheet}
\end{equation}
where $\varepsilon(z) = \text{signum}(z)$.  By concatenation, we recover
\begin{equation}
{\mathbf E}(x) = \int d\tau \; {\mathbf e}(x,\tau) 
= \int d\tau \; {\frac{\sigma }{{2}}}\varepsilon(z)\varphi \left( \tau -t+\frac{\left\vert
z\right\vert }{c}\right) \; \mathbf {\hat{z}}
= {\frac{\sigma }{{2}}}\; \varepsilon(z) \; \mathbf {\hat{z}}
\end{equation}
in agreement with the Maxwell field from a charged sheet.  We notice that as expected, the space
part of the electric
fields change sign at the plane of the sheet, pointing out at each side.
Consequently, an event passing through a charged sheet of equal sign will decelerate in space on
its approach and then accelerate as it retreats.  
On the other hand, unlike the field of a point event,
the temporal part $\epsilon^0$ is an even function of spatial distance
and so the event will accelerate along the time axis 
on both its approach to the charged sheet and its retreat.

\section{Potential Barrier}

We consider the field produced by an infinite charge sheet in the $x-y$ plane at $z=0$,
for which $\sigma e >0 $ leads to a repulsive force on an event of charge $e$.  
Regarding this field as an external force provides a laboratory in which to study the behavior of
an event approaching a potential barrier.  Substituting (\ref{ep0-sheet}) and (\ref{e-sheet}) into 
(\ref{lf}) the Lorentz force on this event is found to be
\begin{eqnarray}
M~\ddot{t}&=&\frac{\lambda e}{c^{2}} \; \mathbf{e}\cdot \mathbf{\dot{x}}+\frac{\lambda e}{c } \; \epsilon ^{0} 
= \frac{\lambda e\sigma}{c }\left[\frac{\varepsilon \left( z\right)}{2c} \;
 \mathbf{\hat{z}}\cdot \mathbf{\dot{x}}+ 1\; \right]
\varphi \left( \tau -t+\dfrac{ \left\vert z\right\vert }{c}\right) 
\\ 
M~\mathbf{\ddot{x}}&=&\lambda e \; \mathbf{e} \; \dot{t}+\frac{\lambda e}{c}
\; \mathbf{\dot{x}} \; {\mathbf \times} \; \mathbf{b}+\lambda e \; \boldsymbol{\epsilon }
= \frac{\lambda e\sigma}{2 }\left[\varepsilon \left( z\right)
 \dot{t}+ 1\; \right]
\varphi \left( \tau -t+\dfrac{ \left\vert z\right\vert }{c}\right) 
\end{eqnarray}
which can be written
\begin{equation}
\dfrac{d}{d\tau }\left[ 
\begin{array}{c}
c\dot{t} \\ 
\dot{z}
\end{array}
\right] = \left[ 
\begin{array}{c}
2\Omega  \left( \dfrac{1}{2}\varepsilon \left( z\right) 
\dfrac{\dot{z}}{c}+1\right)  \rule[-16pt]{0cm}{16pt}  \\ 
\Omega \; \varepsilon \left( z\right)  \left(\dot{t} + 1 \right)
\end{array}
\right]
=\frac{\varepsilon \left( z\right)}{c}\left[ 
\begin{array}{cc}
0 & \Omega \\ 
\Omega & 0
\end{array}
\right] \left[ 
\begin{array}{c}
c\dot{t} \\ 
\dot{z}
\end{array}
\right] +\left[ 
\begin{array}{c}
2\Omega \\ 
\Omega
\end{array}
\right] 
\label{motion}
\end{equation}
when the smoothing function $\varphi (\tau)$ is expressed in terms of the shape $P_{\lambda }$ 
of the barrier potential 
\begin{equation}
\varphi \left( \tau \right)  =\frac{1}{ 2\lambda }P_{\lambda }\left( \tau \right)
\end{equation}
and is given by
\begin{equation}
\Omega(x,\tau) = {\dfrac{e\sigma}{4M}} P_{\lambda }\left( \tau -t+\dfrac{ \left\vert z\right\vert
}{c}\right) .
\end{equation}
%


As an approximation to the smooth function given in (\ref{fi-cur}) we begin with 
a rectangular barrier defined as 
\begin{equation}
P_{\lambda }\left( \tau \right)  =\theta \left( \tau +\lambda \right) -\theta \left( \tau -\lambda \right)
= \left\{
\begin{array}{ll}
1,&  -\lambda \le \tau  \le \lambda \vspace{4pt}\\ 
0,&\text{otherwise} \\ 
\end{array}
\right.
\end{equation}
as the shape function.
The approaching event experiences force wherever
\begin{equation}
-\lambda \le \tau -t+\dfrac{ \left\vert z\right\vert }{c} \le \lambda 
\label{cond}
\end{equation}
leading to the conditions
\begin{eqnarray}
t-\tau &\geq &\frac{\left\vert z\right\vert }{c}-\lambda 
\label{c1}
\\
t-\tau &\leq &\frac{\left\vert z\right\vert }{c}+\lambda 
\label{c2}
\end{eqnarray}
which must be satisfied simultaneously for the barrier to affect the event.
We consider an event approaching the barrier from the right with uniform velocity $u$ so that
\begin{equation}
z(\tau)=Z_{0}-u\tau \mbox{\qquad}Z_{0},u>0 \qquad Z_{0}> \lambda c.
\end{equation}
%
%
%
With these conditions, (\ref{c1}) becomes
\begin{equation}
t-\tau = \left( u^0-1 \right)\tau \geq \frac{\left\vert z\right\vert }{c}-\lambda =\frac{Z_{0}-u\tau }{c}-\lambda
\end{equation}
so that the event reaches the potential barrier when
\begin{equation}
\tau = \tau_0 = \frac{Z_{0}-\lambda c}{u+c\left( u^0 -1\right)} 
\end{equation}
and condition (\ref{c2}) is automatically satisfied.
We can solve (\ref{motion}) as
\begin{equation}
\left[ 
\begin{array}{c}
c\dot{t} (\tau) \rule[-08pt]{0cm}{08pt}\\ 
\dot{z}(\tau) 
\end{array}
\right] 
=\left[ 
\begin{array}{cc}
\cosh \frac{\Omega}{c} \; (\tau - \tau_0)  & \sinh \frac{\Omega}{c} \; (\tau - \tau_0)  \rule[-08pt]{0cm}{08pt}\\ 
\sinh \frac{\Omega}{c} \; (\tau - \tau_0)  & \cosh \frac{\Omega}{c} \; (\tau - \tau_0)  
\end{array}
\right] \left[ 
\begin{array}{c}
c\dot{t} (\tau_0)\rule[-08pt]{0cm}{08pt}\\ 
-u
\end{array}
\right] +\left[ 
\begin{array}{c}
2\Omega \rule[-08pt]{0cm}{08pt}\\ 
\Omega
\end{array}
\right] (\tau - \tau_0)
\vspace{8pt}
\end{equation}
from $\tau = \tau_0$ until the event reaches the charged sheet or reverses direction.


For low energy scattering with initial conditions
\begin{equation}
t(0) = \tau \mbox{\qquad\quad}
\dot{t}(0) = 1
\mbox{\qquad\quad} -u = \dot{z}(0)  = \dfrac{dz}{dt} 
\mbox{\qquad\quad} \vert \dot{z}(0) \vert  = \vert u \vert  \ll c
\end{equation}
%
we find from condition (\ref{c1}) that
the approaching event experiences the potential barrier for times
\begin{equation}
\lambda \geq \frac{\left\vert z\right\vert }{c}= \frac{
Z_{0}-u\tau }{c} \ \longrightarrow \ \tau \geq \tau_0 = \frac{Z_{0}-\lambda c}{u} 
\end{equation}
and notice that condition (\ref{c2}) is satisfied automatically because $t \simeq \tau$.
The equations of motion (\ref{motion}) reduce to
\begin{equation}
~\ddot{t} \simeq{\frac{e\sigma}{2Mc}}P_{\lambda }\left( \frac{ \left\vert z\right\vert }{c}\right)
\mbox{\qquad\quad} 
~\ddot{z} \simeq{\frac{e\sigma}{2M}}\varepsilon \left( z\right) P_{\lambda
}\left( \frac{\left\vert z\right\vert }{c}\right) 
\end{equation}
so that the event will decelerate along the $z$-axis until it either reverses direction or passes through
the charged sheet.
For short times, such that
${\frac{e\sigma}{2Mc}} (\tau - \tau_0) < 1$, 
we may write the solutions
\begin{equation}
\dot{t}(\tau)  \simeq 1  + {\dfrac{e\sigma}{2Mc}}\; (\tau - \tau_0)
\mbox{\qquad\quad} \dot{z}(\tau) \simeq -u + {\dfrac{e\sigma}{2M}} (\tau - \tau_0) 
\end{equation}
from which we notice that despite the spatial deceleration, the event accelerates in time. 
To compare this result with the energy
of a particle evolving on its mass-shell we calculate the $t$-velocity
\begin{equation}
\dot{t}_\text{on-shell} = \dfrac{1}{ \sqrt{1 - \dfrac{\dot{z}^2}{c^2} } } = 
\dfrac{1}{ \sqrt{1 - \left[ {\dfrac{e\sigma}{2Mc}} (\tau - \tau_0) - \dfrac{u }{c} \right]^2 } }
\simeq 1 + \dfrac{1}{2}\left[ {\dfrac{e\sigma}{2Mc}} (\tau - \tau_0) - \dfrac{u }{c} \right]^2
\end{equation}
so that the $t$-acceleration
\begin{equation}
\ddot{t}_\text{on-shell}  \simeq 
-  \dfrac{e\sigma}{2Mc} \left[\dfrac{u }{c}  - {\dfrac{e\sigma}{2Mc}} (\tau - \tau_0) \right] < 0
\end{equation}
is negative as expected.  The $t$-acceleration of the off-shell event corresponds to a
transfer of mass from the field to the event found from (\ref{mass}) to be
\begin{equation}
\frac{d}{d\tau }\left( -\frac{1}{2c^2}M\dot{x}^{2} \right) =-\frac{e_{0}}{c^2}\ \dot{x }^{\mu }f_{\mu 5} 
\simeq \frac{e_{0}}{c}\; \dot{t } \; \epsilon^0 
\simeq \dfrac{e\sigma}{2c} \; P_\lambda .
\end{equation}
If the charge density is low enough to permit the event to pass through the charge sheet, then 
neglecting the effect of deceleration on the transit time,
the total mass shift of the event is on the order of $\lambda e \sigma / 2u $.


As an interesting example of 5D electrostatics, we consider a test event at rest within the 
potential barrier at time $\tau = 0$
\begin{equation}
0<z\left( 0\right) =Z_{0}<\lambda c\mbox{\qquad}\mbox{\quad} \dot{z}\left( 0\right) = 0 
\mbox{\qquad} \mbox{\quad}\dot{t}(0) = 1 
\mbox{\qquad} \mbox{\quad}t(0) = 0
\end{equation}
and bound in an insulator so that
\begin{equation}
\ddot{t} \simeq {\dfrac{e\sigma}{2Mc}}P_{\lambda }\left( \tau -t+\frac{Z_{0}}{c}\right) \mbox{\qquad} \mbox{\qquad} 
\ddot{z} = \frac{F_{\text{electric}} -F_{\text{binding}}}{M}=0 .
\end{equation}
In this case, condition (\ref{c1}) is automatically satisfied
so we may write
\begin{equation}
t =  \left\{
\begin{array}{ll}
\tau + \dfrac{1}{2} \; {\dfrac{e\sigma}{2Mc}} \tau^2, \quad&  0 \leq \tau \leq \tau_c\vspace{4pt}\\ 
t(\tau_c) + \left[ 1 + {\dfrac{e\sigma}{2Mc}} \right] (\tau - \tau_c) ,&\tau > \tau_c \\ 
\end{array}
\right.
\end{equation}
%
where, by condition (\ref{c2}),  
\begin{equation}
\mbox{\qquad} t-\tau = \dfrac{Z_{0}}{c}+\lambda \mbox{\qquad} \Rightarrow \mbox{\qquad}
\tau _{c} =\sqrt{\frac{4M\left( Z_{0}+\lambda c\right) }{e\sigma}} .
\end{equation}
Thus, the event experiences a short-lived acceleration which pushes it beyond the range of the
potential barrier at $\tau_c$.
For $\tau > \tau_c$, the temporal velocity remains constant at
\begin{equation}
\dot{t}\left(\tau_{c}\right) = 1 + {\dfrac{e\sigma}{2Mc}} \tau _{c}
=1+\sqrt{\frac{\left( Z_{0}+\lambda c\right)e\sigma }{Mc^2}}
\mbox{\quad} 
\end{equation}
so that during the interaction with the potential barrier, the squared mass of the event grows from
$M^2$ to
\begin{equation}
m^2\left(\tau_c\right) = M^2 \left( \dfrac{\dot{x}}{c} \right)^2  = M^2 \dot{t}^2 
= M^2\left[ 1+\sqrt{\frac{\left( Z_{0}+\lambda c\right)e\sigma }{Mc^2}}\right]^2.
%
\label{191}
\end{equation}
However, if the event has opposite charge $e=-\vert e\vert$ then the event will decelerate in time as
\begin{equation}
\ddot{t} \simeq - {\dfrac{\vert e\vert\sigma}{2Mc}}P_{\lambda }\left( \tau -t+\frac{Z_{0}}{c}\right) \mbox{\qquad} \mbox{\qquad} 
\ddot{z} = \frac{F_{\text{electric}} -F_{\text{binding}}}{M}=0 
\end{equation}
and condition (\ref{c1}) leads to
\begin{equation}
\dot{t}\left(\tau\right) = 1 - {\dfrac{\vert e\vert \sigma}{2Mc}} \tau
\end{equation}
suggesting that for a sufficient charge density the event could reverse direction in time.
By condition (\ref{c1}) the time evolution is
\begin{equation}
t =  \left\{
\begin{array}{ll}
\tau - \dfrac{1}{2} \; {\dfrac{\vert e\vert\sigma}{2Mc}} \tau^2, \quad& 
0 \leq \tau \leq \tau_c\vspace{4pt}\\ 
t(\tau_c) + \left[ 1 - {\dfrac{e\sigma}{2Mc}} \right] (\tau - \tau_c) ,&\tau > \tau_c \\ 
\end{array}
\right.
\qquad \qquad \tau _{c} =\sqrt{\frac{4M\left( \lambda c - Z_{0}\right) }{e\sigma}}  
\end{equation}
so that for $\tau > \tau_c$ the time evolution is 
\begin{equation}
\dot{t}\left(\tau_{c}\right) = 1 - {\dfrac{e\sigma}{2Mc}} \tau _{c}
=1-\sqrt{\frac{\left( \lambda c - Z_{0}\right)e\sigma }{Mc^2}}
\label{183}
\end{equation}
and
\begin{equation}
m^2\left(\tau \geq \tau_c\right)
= M^2\left[ 1-\sqrt{\frac{\left( \lambda c - Z_{0}\right)e\sigma }{Mc^2}}\right]^2
%
\label{184}
\end{equation}
is the squared mass.  In the case that the event emerges from the
interaction with the field as an anti-event with $\dot{t}\left(\tau_{c}\right) = -1$, it 
emerges with its initial mass.


For the case of an event held at rest in space, we are not restricted to the approximation of the
rectangular barrier and can find a solution for the smooth potential given by
\begin{equation}
\varphi\left(\tau\right) = \dfrac{1}{2\lambda}e^{-\left\vert \tau \right\vert /\lambda }
\mbox{\quad} \rightarrow \mbox{\quad} P_\lambda \left(\tau\right)
= e^{-\left\vert T_{0}-\left( t-\tau \right) \right\vert /\lambda } \qquad \qquad T_0 =
\dfrac{Z_0}{c} .
\end{equation}
The equations of motions are now nonlinear
\begin{equation}
\ddot{t} \simeq {\dfrac{e\sigma}{2Mc}}e^{-\left\vert T_{0}-\left( t-\tau \right) \right\vert /\lambda }
\mbox{\qquad} \mbox{\qquad} 
\ddot{z} = \frac{F_{\text{electric}} -F_{\text{binding}}}{M}=0 
\end{equation}
with the initial conditions $z(0) = Z_0$ and $t(0) = \tau = 0$.
As the event accelerates in this field, the $t$ will grow larger than $\tau$ and the field strength will
increase to a maximum when $t-\tau = T_{0}$ and then decrease when $t-\tau > T_{0}$.  Introducing
the variables
\begin{equation}
s_1 = t-\tau < T_{0} \mbox{\qquad} \mbox{\qquad}
-s_2 = t-\tau > T_{0}
\end{equation}
the approach to the field maximum is described by 
\begin{equation}
\ddot{s}_1={\dfrac{e\sigma}{2Mc}}e^{-T_{0}/\lambda } \; e^{s_1/ \lambda}={\alpha } \; e^{s_1/\lambda }
\qquad \qquad 
\alpha  = {\dfrac{e\sigma}{2Mc}}e^{-T_{0}/\lambda }
\label{approach}
\end{equation}
and the retreat from the field maximum is described by
\begin{equation}
\ddot{s}_2=-{\dfrac{e\sigma}{2Mc}}e^{T_{0}/\lambda } \; e^{s_2/ \lambda}=- {\beta } \;
e^{s_2/\lambda } 
\qquad \qquad 
\beta  = {\dfrac{e\sigma}{2Mc}}e^{T_{0}/\lambda } \; .
\label{retreat}
\end{equation}
The generic equations (\ref{approach}) and (\ref{retreat}) may be solved by writing
\begin{equation}
\ddot{s}_1={\alpha } \; e^{s_1/\lambda }\mbox{\quad} \rightarrow \mbox{\quad} 
\dot{s}_1 \; \ddot{s}_1=\frac{d}{d\tau }\left[\frac{1}{2}\left( \dot{s}_1\right) ^{2}\right]  \ = \
\dot{s}_1 \; {\alpha }e^{s_1/\lambda }
=\frac{d}{d\tau }\left[\lambda {\alpha }e^{s_1/\lambda }\right]
\label{nl-solve}
\end{equation}
so that $\tau$-integration leads to 
\begin{equation}
\dot{s}_1 (\tau) = \sqrt{2{\alpha }\lambda }\sqrt{e^{ s_1/\lambda}-1}
\end{equation}
where we used $\dot{s}_1 (0) = \dot{t} - 1 = 0 $.  Now integrating
\begin{equation}
\int \frac{1}{\sqrt{e^{{ s_1/\lambda}}-1}} \; \frac{ds_1}{d\tau } d\tau = 2\lambda \tan ^{-1}\sqrt{e^{ s_1/\lambda }-1}
= \sqrt{2{\alpha }\lambda } \tau +c_1
\end{equation}
and using $s_1 (0) = t(0) - 0 = 0 $ we find
\begin{equation}
s_1 =\lambda \log \left( \tan ^{2}\left( \sqrt{\frac{{\alpha }}{2\lambda }} \tau \right) +1\right) 
\label{193}
\end{equation}
which becomes
\begin{equation}
t =\lambda \log \left( \tan ^{2}\left( \sqrt{\frac{{\alpha }}{2\lambda }}\tau \right) +1\right)
+\tau . 
\end{equation}
Designating $\tau_0$ as the time when the field reaches its maximum, 
so that $s_1 \left( \tau_0 \right) = T_0$, we may invert (\ref{193}) to find
\begin{equation}
\tau _{0} =\sqrt{2\lambda /\alpha }\;\tan ^{-1} \sqrt{e^{T_0/\lambda}-1}
\end{equation}
and
\begin{equation}
\dot{t}\left( \tau _{0}\right)  = 1 + \sqrt{ \frac{\lambda e\sigma }{Mc} \left( 1-e^{-T_0 / \lambda
}\right)} .
\end{equation}

For $\tau >\tau _{0} $ we apply this generic solution to (\ref{retreat}) using
initial conditions
\begin{equation}
s_2\left(\tau_0\right) = T_0 \qquad \qquad
\dot{s}_2\left( \tau _{0}\right) =
1-\dot{t}\left( \tau _{0}\right) = - \sqrt{2{ \alpha }\lambda \left(e^{T_{0}/\lambda }-1\right)}
\label{198}
\end{equation}
to find
\begin{equation}
\dot{s}_2=-\sqrt{2{\beta }\lambda }\sqrt{C-e^{s_2/\lambda}}  
\end{equation}
with constant of integration 
\begin{equation}
C=e^{- 2T_0/\lambda}\left( 2e^{T_0/\lambda}-1\right)  .
\end{equation}
Integrating again
\begin{equation}
\int \frac{1}{\sqrt{C-e^{s_2/\lambda}}} \; \frac{ds_2}{d\tau } d\tau
= -\dfrac{2\lambda}{C} \tanh ^{-1}\sqrt{\dfrac{C-e^{s_2/\lambda}}{C}}
= - \sqrt{2{\beta }\lambda } \tau +C
\end{equation}
we arrive at
\begin{equation}
s_2 = \lambda \log \left[ C-C\tanh ^{2}\left( \sqrt{\frac{\beta C}{2\lambda}}\tau +
\sqrt{2{\beta }\lambda }\tau _{0}- \frac{2\lambda}{\sqrt{C}}
\tanh ^{-1}\sqrt{\frac{e^{T_0/\lambda}-1}{2e^{T_0/\lambda}-1}}\right) \right]
\label{200}
\end{equation}
%
which becomes
\begin{equation}
t = \tau + \lambda \log \left[\dfrac{1}{C}\cosh ^{2}\left( \sqrt{\frac{\beta C}{2\lambda}}\tau +
\sqrt{2{\beta }\lambda }\tau _{0}- \frac{2\lambda}{\sqrt{C}}
\tanh ^{-1}\sqrt{\frac{e^{T_0/\lambda}-1}{2e^{T_0/\lambda}-1}}\right) \right] . 
\end{equation}
Inserting (\ref{200}) into (\ref{198}) we find
\begin{equation}
\dot{s}_2= 1- \dot{t} = -\sqrt{2{\beta }\lambda C}\tanh \left( \sqrt{\frac{{\beta C}}{2\lambda }}\tau +
\sqrt{2{\beta }\lambda }\tau _{0}-\frac{2\lambda }{\sqrt{C}}\tanh ^{-1}
\sqrt{\frac{e^{T_0/\lambda}-1}{2e^{T_0/\lambda}-1}}
\right) 
\end{equation}
and 
\begin{equation}
\dot{t} = 1 + \sqrt{2{\beta }\lambda C}\tanh \left( \sqrt{\frac{{\beta C}}{2\lambda }}\tau +
\sqrt{2{\beta }\lambda }\tau _{0}-\frac{2\lambda }{\sqrt{C}}\tanh ^{-1}
\sqrt{\frac{e^{T_0/\lambda}-1}{2e^{T_0/\lambda}-1}}
\right) .
\end{equation}
To determine the asymptotic mass of the event we calculate
\begin{equation}
\dot{t} \underset{\tau \rightarrow \infty}{\mbox{\quad}\xrightarrow{\hspace*{1cm}}\mbox{\quad}}
 \dot{t}_{\max } = 1 + 
\sqrt{{\frac{\lambda e \sigma}{Mc}}}\sqrt{2-e^{-T_{0}/\lambda} }
\end{equation}
so that
\begin{equation}
m^{2}\longrightarrow M^{2}\left( 1+\sqrt{ {\frac{\lambda e\sigma}{Mc}}}\sqrt{2-e^{-\frac{Z_{0}}{\lambda c}}}\right)
^{2}\simeq M^{2}\left( 1+\sqrt{{\frac{ (Z_0 +\lambda c )e\sigma}{Mc^{2}}}} \right)^{2} 
\end{equation}
which is identical to (\ref{191}) in the low energy approximation.
As is previous examples, the field transfers mass to the event, despite the deceleration in space.

\section{Discussion}

In Stueckelberg-Horwitz electrodynamics, particle worldlines are traced out through the evolution
of interacting spacetime events $x^\mu(\tau)$ whose associated current densities induce the
$\tau$-dependent fields that mediate their interactions.  By introducing the chronological
time $\tau$ as an independent evolution parameter, and freeing the laboratory clock $x^0$ to 
propagate alternately 'forward' and 'backward' in time according to the sign of its energy,
this formalism provides a classical implementation of the Feynman-Stueckelberg interpretation
of pair creation/annihilation.
However, as Stueckelberg discovered, allowing $\dot x^0$ to evolve independently of $\dot{{\mathbf x}}$ 
is not sufficient to permit classical pair creation/annihilation because mass is
conserved for interactions mediated by the antisymmetric electromagnetic tensor.  To evolve outside
the forward or reverse timelike region an event must undergo an exchange of mass with an external
field.

It was shown in \cite{beyond} that (\ref{lag}) is the most general classical Lagrangian consistent with 
the quantum commutation relations
\begin{equation}
[x^\mu,x^\nu] = 0 \qquad \qquad M [x^\mu,\dot x^\nu] = -i\hbar \eta^{\mu\nu}
\end{equation}
among unconstrained phase space variables.  Thus, Stueckelberg-Horwitz electrodynamics follows
from two assumptions: a kinetic term of the type (\ref{field-kin}) for the fields modeled on Maxwell
theory, and a phase space on which the usual mass-shell constraint $p^\mu p_\mu =-m^2$
is relaxed.  By demoting mass conservation from an {\em a priori} constraint to a conservation law
that applies for appropriate interactions, the formalism acquires its key features: 
integrability, local gauge symmetry with five $\tau$-dependent gauge fields, classical
implementation of negative energy evolution, and exchange of mass between particles and fields.  

In a Lorentz-covariant generalization of the nonrelativistic picture, event evolution is associated with 
a current consisting of a scalar event density
$j^5(x,\tau)$ that characterizes the event distribution in spacetime, and a vector current $j^\mu(x,\tau)$
that characterizes the motion of events through spacetime. In this picture, current conservation
describes the 4-divergence of the spacetime current balancing the $\tau$-variation of the
event density.  
Even a 'static' event must move along its time axis, and so both the scalar
current $j^5(x,\tau)$ and the temporal component $j^0(x,\tau)$ of the vector current contribute to gauge fields
$a^5(x,\tau)$ and $a^0(x,\tau)$ that produce field strengths $e^i = f^{0i}$ and $\epsilon^\mu = f^{5\mu}$.
As seen in the structure of the pre-Maxwell equations, the $\epsilon^\mu$
field plays a role analogous to the Maxwell electric field --- its divergence points to the scalar 
source inducing the field.  Similarly, the field $f^{\mu\nu}$ plays a role analogous to the Maxwell magnetic
field --- it has no monopoles and is induced by the motion of the source through spacetime.

In this paper we studied simple solutions for timelike particles in uniform motion.  
Calculating the fields induced by these events, we confirmed that they satisfy the pre-Maxwell
field equations.  We found the $\tau$-dependent potentials for the
moving event and the field strengths $e^i = f^{0i}$ and $\epsilon^\mu = f^{5\mu}$ which take on a generalized 
Coulomb-like form.  For a test event at rest with respect to the source event, a time
acceleration was observed suggesting a transfer of mass between the field and test event.  
For this solution, the integral forms of Gauss's law and Faraday's law were
found by extending the integrations to 4D volumes of spacetime.  As in Maxwell theory,
the evaluation of Stoke's theorem shows that the line integral of the electric field $\epsilon^\mu$
around a closed path vanishes for paths in space or long paths in spacetime.  However, it was seen
that the smoothed structure of the current suggests that a net change in mass may result from the
motion of an event around a short closed timelike path.  These effects were the first of several
indications of mass changing effects in 5D electrostatics.
The fields produced by a line charge and a charged sheet were also calculated.  The fields for the
three source configurations were shown to recover the standard Maxwell fields under concatenation.

The field induced by the infinite charged sheet was used to study the evolution of a test charge
approaching a potential barrier.  It was seen that a low-energy event approaching the barrier would
undergo a  temporal acceleration (energy growth) while the event decelerated away from the barrier.  We
also studied the evolution of a test event held fixed in space in an insulator, and saw that from
the onset of the configuration, the field would transfer mass to the test event as it gained energy
(accelerates in $t$) with no corresponding change in spatial position or velocity.  This effect was
confirmed for both a rectangular barrier and the exponential shape associated with the smoothing
function.  Intriguingly, in these mass shift effects, the period of mass change is short-lived,
because the field pushes the test event out of range, and so the total mass shift is small and
finite.  These mass shift effects suggest several questions related to the observed conservation
of elementary particle mass.  It would be interesting to find that just as
a regular structure of particles remains at rest in an insulator, so 
the net effect of the $\epsilon^\mu$ field in a regular structure produces a mass
insulation effect.  This speculation will be addressed in a forthcoming paper.



%
%
%

%
%
%

\begin{thebibliography}{99}
%
\bibitem{B-W_p.xxv-xxvi} Born and Wolf 1975 {\em Principles of Optics} fifth edition (Oxford: Pergamon Press) xxv-xxvi.
\bibitem{fock_gauge} For discussion of the conceptual history of gauge theory see J. D. Jackson 2001 Rev. Mod. Phys. {\bf 73} 663.
\bibitem{Stueckelberg} Stueckelberg E C G 1941 \textit{Helv. Phys. Acta}
\textbf{14}  322;
Stueckelberg E C G 1941 \textit{Helv. Phys. Acta} \textbf{14}  588
\bibitem{Feynman} Feynman R P 1950 \textit{Phys. Rev.} \textbf{80} 440;
Feynman R 1948 \textit{Rev P. Mod. Phys.}  \textbf{20} 367 
\bibitem{Schwinger} Schwinger J 1951 \textit{Phys.\ Rev.} \textbf{82}  664
\bibitem{H-P} Horwitz L P and Piron C 1973 \textit{Helv. Phys. Acta} \textbf{48} 316 
\bibitem{saad} Saad D, Horwitz L P and Arshansky R I 1989 \textit{ Found.\ of Phys.} \textbf{19}  1126
\bibitem{beyond} Land M C, Shnerb N and Horwitz L P 1995 \textit{J.\ Math.\ Phys.} \textbf{36}  3263
\bibitem{emlf} Land M C and Horwitz L P 1991 \textit{Found. of Phys. Lett.} \textbf{4}  61
\bibitem{I} Arshansky R I and Horwitz L P 1989 \textit{J.\ Math.\ Phys.} \textbf{30}  66
\bibitem{II} Arshansky R I and Horwitz L P 1989 \textit{J.\ Math.\ Phys.} \textbf{30}  380
\bibitem{scattering} Horwitz L P and Lavie Y 1982 \textit{Phys.\ Rev.} D
\textbf{26}  819;
Arshansky R I and Horwitz L P 1989 \textit{J.\ Math.\ Phys.} \textbf{30}
213;
Arshansky R I and Horwitz L P 1988 \textit{Phys.\ Lett} A \textbf{131}  222
\bibitem{zeeman} Land M C and Horwitz L P 1995 \textit{J.\ Phys.\ } A \textbf{28}  3289
\bibitem{stark} Land M C 2001 \textit{Found.\ of Phys.} \textbf{31}  967
\bibitem{larry} Land M C 1996 \textit{Found.\ of Phys.} \textbf{27}  19
\bibitem{letter}  Land M C and Horwitz L P 1998 Land M C  \textbf{A239} 135
\bibitem{ald}  Land M C 2011 \textit{J.\ Phys.: Conf.\ Ser.} \textbf{330} 012015
\bibitem{concat} Arshansky R L, Horwitz P and Lavie Y 1983 \textit{ Found.\ of Phys.} \textbf{13}  1167
\bibitem{green} Land M C and Horwitz L P 1991 \textit{Found.\ of Phys.} \textbf{21}  299
\bibitem{jigal} Aharonovich I and Horwitz L P 2012 \textit{J.\ Math.\ Phys.} {\bf 53} 032902
\bibitem{pdg} Yao W-M et.\ al.\ 2006 (Particle Data Group) \textit{J.\ Phys.\ } G \textbf{33} 1
\bibitem{high-order} Land M C 2003 \textit{Found.\ of Phys.} \textbf{33}  1157
%
\end{thebibliography}
\end{document}